\documentstyle[cjaa209,twoside,epsf]{article}
\input{psfig.sty}

\begin{document}

\title{ Non-LTE Analysis for the Sodium Abundances of Metal-Poor Stars
in the Galactic Disk and Halo}

\author{Yoichi Takeda$^{1}$\mailto{}, Gang Zhao$^{2}$, Masahide 
Takada-Hidai$^{3}$, Yu-Qin Chen$^{2}$, Yu-ji Saito$^{4}$, and 
Hua-Wei Zhang$^{5,6}$}

\inst{
$^1$ Komazawa University, Komazawa, Setagaya,Tokyo 154-8525, Japan\\
$^2$ National Astronomical Observatories, Chinese Academy of Sciences,
              Beijing 100012 \\
$^3$ Liberal Arts Education Center, Tokai University, Hiratsuka, Kanagawa 259-1292, Japan\\
$^4$ Department of Physics, Faculty of Science, Tokai University, Hiratsuka, Kanagawa 259-1292, Japan\\
$^5$ Department of Astronomy, School of Physics, Peking University, Beijing 100871\\
$^6$ CAS-Peking University Joint Beijing Astrophysical Center, Beijing 100871\\ 
}

\email{takedayi@cc.nao.ac.jp}

\markboth{Y. Takeda, G. Zhao, M. Takada-Hidai, et al.}
{Non-LTE Analysis for the Sodium Abundances of Metal-Poor Stars in
the Galactic Disk and Halo}

\pagestyle{myheadings}

\date{Received 2003 March 10; accepted 2003 April 18}

\baselineskip=18pt

\begin{abstract}
\baselineskip=18pt
We performed an extensive non-LTE analysis of
the neutral sodium lines of Na~{\sc i} 5683/5688, 5890/5896,
6154/6161, and 8183/8195 for disk/halo stars of F--K type covering
a wide metallicity range ($-4 \la$~[Fe/H]~$\la 0.4$), based
on our own data as well as those collected from the literature.
For comparatively metal-rich disk stars ($-1
\la$~[Fe/H]~$\la 0$) where the weaker 6154/6161 lines are
best abundance indicators, we confirmed [Na/Fe]~$\sim$~0 with an
``upturn'' (i.e., a shallow/broad dip around $-0.5 \la$ [Fe/H]
$\la 0$) as already reported by previous studies. Regarding the
metal-deficient halo stars, where the much stronger 5890/5896 or
8183/8195 lines suffering considerable (negative) non-LTE
corrections amounting to 0.5~dex have to be used, our analysis
suggests mildly ``subsolar'' [Na/Fe] values down to $\sim -0.4$
(with a somewhat large scatter of $\sim \pm 0.2$~dex) on the average
at the typical halo metallicity of [Fe/H]~$\sim -2$, while they
appear to rise again toward a very metal-poor regime recovering a
near-solar ratio of [Na/Fe]~$\sim$~0 ([Fe/H] $\sim -3$ to $-4$).
These results are discussed in comparison with the previous
observational studies along with the theoretical predictions from
the available chemical evolution models.
\end{abstract}

\keywords{Galaxy: evolution --- line: formation --- 
stars: abundances --- stars: atmospheres --- stars: late-type}

\section{Introduction}
\label{sect:intro} The purpose of this paper is to clarify the
behaviors of the abundances of sodium in disk/halo stars in our
Galaxy.

Recently, we carried out an extensive non-LTE abundance analysis
of potassium based on the K~{\sc i} 7699 resonance line
and found a tendency of mildly supersolar [K/Fe] ratios for
metal-poor stars just similar to the case of $\alpha$-capture
elements (Takeda et al. 2002). As a continuation of that study,
we pay attention this time to another similar alkali element, 
sodium (Na).

The main site of Na synthesis is considered to be
the hydrostatic carbon burning inside massive stars
(e.g., Woosley, Weaver 1995), though other manufacturing processes
(e.g., NeNa cycle in the H-burning shell, type I supernova, etc.)
are possible and may have an appreciable affection on its chemical
history (see, e.g., Timmes et al. 1995).
As an odd-$Z$ (neutron-rich) element, its production is sensitive
to the neutron excess and therefore metal-dependent.
Hence, clarifying the behavior of its abundance in terms of
the cosmic metallicity would provide us important
information concerning the variation of the yield, the detailed
mechanism of the synthesis, and the Galaxy evolution models
(initial mass function, infall, etc ).

Unlike the case of potassium, a reasonable number of adequate lines
are available for sodium abundance determinations, which lie mostly
in the visible through near-IR wavelength regions. Hence, many observational
studies have so far been performed over these two decades
(see, e.g., the references quoted in the recent theoretical papers
such as Timmes et al. 1995, Samland 1998, and Goswami, Prantzos 2000).
Nevertheless, the behavior of [Na/Fe] is currently still controversial
and has not yet been settled.

It has generally been considered until mid-1990's that Na almost
scales with Fe at all metallicities though with a considerable scatter
around [Na/Fe]~$\sim$~0. [See, for example, subsubsection 3.5.3 of
Timmes et al. (1995) for a review on the situation at the time of
early 1990's.] Actually, a nearly solar ratio was obtained
even for extremely metal-poor stars ([Fe/H] down to $\sim -4$)
by the extensive study of McWilliam et al. (1995b) using the
resonance D lines (Na~{\sc i} 5890/5896). It should be pointed out
that most of these previous studies were done based on the assumption
of LTE.

In late 1990's, however, new studies have appeared that cast doubt
to this classical picture. Baum{\" u}ller et al. (1998) investigated
the non-LTE formation of various Na~{\sc i} lines and found considerable
(negative) non-LTE corrections reaching $-0.5$ dex for the D lines,
by which a systematic tendency of {\it decreasing} [Na/Fe] ratios
is obtained toward a metal-poor regime ([Na/Fe] $\sim -0.5$ at
[Fe/H] $\sim -3$). Similarly, the (LTE) analysis of Stephens (1999)
for halo dwarf stars based on high-quality spectra of Na~{\sc i}
5683/5688 subordinate lines suggested that [Na/Fe] progressively
decreases (from [Fe/H] $\sim -1$ to [Fe/H] $\sim -2$) down to
the value of $\sim -0.5$.

Accordingly, it appears that a proper account of the non-LTE effect
is mandatory for studying stellar sodium abundances, at least for
the resonance Na~{\sc i} 5890/5896 D lines visible even in extremely
metal-poor stars. Then, what about other subordinate lines, such as
Na~{\sc i} 6154/6161 lines (widely used for relatively metal-rich
population I disk stars), 5683/5688 lines (often used for typical halo 
stars with [Fe/H] of $-1$ to $-2$), 8183/8195 lines (not so popular due to
their unfavorable wavelength location but potentially useful
abundance indicators because of the clear visibility even down to
[Na/H] $\sim -3$)? Do they also suffer an appreciable non-LTE effect ?
Any definite conclusion on the Galactic [Na/Fe] vs. [Fe/H] behavior
should await until this problem is clarified, because different lines
are used case by case depending on the stellar metallicity.

Admittedly, several elaborate and important studies have recently 
been done concerning the non-LTE effect on Na abundance determinations 
for solar-type stars (e.g., Baum{\" u}ller et al. 1998; Gratton et al. 
1999; Korotin, Mishenina 1999; Mashonkina et al. 2000). We would say,
however, that their results are not presented in sufficient detail
for the purpose of practical applications; i.e., they tend to be
confined to only specific lines of interest or only
representative atmospheric parameters, and it is not necessarily
easy for the reader to assess how all those Na~{\sc i}
lines of importance are affected by the non-LTE effect for stars
showing a diversity of atmospheric parameters.

In view of this necessity, we decided to (1) carry out non-LTE
calculations on the neutral sodium for a wide range of atmospheric
parameters with an intention of applying to those 4 line pairs
mentioned above, (2) construct useful grids of non-LTE
corrections, and (3) perform extensive abundance analysis toward
establishing the Na abundances of disk/halo stars with a variety of
metallicities based on our own new observations/measurements along with
the published equivalent widths taken from various literature.

The following sections of this paper are organized as follows.
Our statistical-equilibrium calculations on neutral sodium
and the resulting non-LTE corrections are explained in section 2,
where we also discuss uncertain factors or adopted approximations
involved in the abundance determination based on some theoretical
test calculations. Our observational data obtained at two
observatories are explained in section 3, followed by section 4 where
we describe the procedures of the abundance analysis using these data
along with the reanalysis of the extensive literature data.
We discuss in section 5 the results obtained in section 4,
especially in terms of the [Na/Fe] vs. [Fe/H] relation and its
implications, while referring to the representative
theoretical calculations of Galactic chemical evolution.
Finally, the conclusion of this paper is presented in section 6.

\section{Non-LTE Calculations}
\label{sect:Non-LTE}
\subsection{Computational Procedures}

The procedures of our statistical-equilibrium calculations for
neutral sodium, based on a Na~{\sc i} atomic model comprising
92 terms and 178 radiative transitions, are essentially the same as
those described in Takeda and Takada-Hidai (1994) and Takeda (1995),
which should be consulted for details.
One difference is the choice of the background model atmospheres;
i.e., instead of Kurucz's (1979) ATLAS6 models adopted in those old
works, the photoionizing radiation field was computed based on
Kurucz's (1993a) ATLAS9 model atmospheres while incorporating
the line opacity with the help of Kurucz's (1993b) opacity
distribution functions.
It should also be mentioned that a reduction factor of 0.1
(logarithmic correction of $h = -1$ according to the definition
of Takeda 1995) was applied to the H~{\sc i} collision rates
computed with classical approximate formula (Steenbock, Holweger 1984;
Takeda 1991) according to the empirical determination of Takeda (1995),
though test calculations by varying this correction factor
from 1 to $10^{-3}$ were also performed (cf. subsection 2.3 below).

Since we planned to make our calculations applicable to stars
from near-solar metallicity (population I) down to very low
metallicity (extreme population II) at late-F through early-K
spectral types in various evolutionary stages
(i.e., dwarfs, subgiants, giants, and supergiants),
we carried out non-LTE calculations on an extensive grid of
125 ($5 \times 5 \times 5$) model atmospheres resulting from
combinations of five $T_{\rm eff}$ values
(4500, 5000, 5500, 6000, 6500 K), five $\log g$ values
(1.0, 2.0, 3.0, 4.0, 5.0), and five metallicities (represented by [Fe/H])
(0.0, $-1.0$, $-2.0$, $-3.0$, $-4.0$).
As for the stellar model atmospheres, we adopted Kurucz's (1993a) ATLAS9
models corresponding to a microturbulent velocity ($\xi$) of 2~km~s$^{-1}$.

Regarding the sodium abundance used as an input value in non-LTE
calculations, we assumed
$\log\epsilon_{\rm Na}^{\rm input}$ = 6.33 + [Fe/H],
where the solar sodium abundance of 6.33 was adopted from
Anders and Grevesse (1989) (which is used also in the ATLAS9 models).
Namely, a metallicity-scaled sodium abundance was assigned to
metal-poor models. The microturbulent velocity (appearing in the
line-opacity calculations along with the abundance) was assumed
to be 2~km~s$^{-1}$, to make it consistent with the model atmosphere.

\subsection{Characteristics of the Non-LTE Effect}

In figure 1 are shown the $S_{\rm L}(\tau)/B(\tau)$ (the ratio of
the line source function to the Planck function, and nearly equal to
$\simeq b_{\rm u}/b_{\rm l}$, where $b_{\rm l}$ and $b_{\rm u}$ are
the non-LTE departure coefficients for the lower and upper levels,
respectively) and  $l_{0}^{\rm NLTE}(\tau)/l_{0}^{\rm LTE}(\tau)$
(the NLTE-to-LTE line-center opacity ratio, and nearly equal to
$\simeq b_{\rm l}$) for each of the multiplet 1 (5890/5896) and
multiplet 4 (8183/8195) transitions (non-LTE effects are especially
important for these two doublets; cf. subsection 5.1)
for a representative set of model atmospheres.
We can read the following characteristics from this figure,
which are mostly the same as those obtained for the case of
K~{\sc i} 7699 (Takeda et al. 2002):\\
--- In almost all cases, the inequality relations of $S_{\rm L}/B<1$
(dilution of line source function) and $l_{0}^{\rm NLTE}/l_{0}^{\rm LTE} >1$
(enhanced line-opacity) hold in the important line-forming region
for both cases of multiplets 1 and 4, which means that the non-LTE
effect almost always acts in the direction of strengthening the 5890/5896 and
8183/8195 lines.\footnote{An exception is the case of lowest
$T_{\rm eff}/\log g$ (e.g., $T_{\rm eff}$=4500~K, $\log g =1.0$, [Fe/H] = 0.0),
where the line can be marginally weakened (i.e., positive abundance
correction) by the enhanced $S_{\rm L}$ over $B$ in the line-forming
region (cf. the footnote in subsection 5.3).}
Actually, more important is the former $S_{\rm L}$-dilution effect
which appreciably deepens/darkens the core of saturated lines in such a way
that would never be accomplished under the assumption of LTE [where the
residual intensity can not become lower than $B(\tau \sim 0)/B(\tau \sim 1)$];
hence, this raises the flat part of the curve of growth, and if this increase
in the saturated-line strength is to be accounted for within the framework 
of LTE, a large abundance variation would have to be invoked. As a result, 
the extent of the non-LTE effect significantly depends on the line-strength 
in the sense that it becomes most conspicuous for the lines in the flat part. 
(See, e.g., St\"urenburg, Holweger 1990, especially figure 15 of 
subsection 4.3 therein.) This situation is demonstrated in
figure 2, where the non-LTE corrections (see subsection 2.3 below)
for the 5896 and 8195 lines are shown as functions of the equivalent
width for representative model atmospheres.\\
--- There is a tendency that the non-LTE effect is enhanced
with a lowering of the gravity, as naturally expected.\\
--- The departure from LTE appears to be larger for higher $T_{\rm eff}$
in the high-metallicity (1$\times$) case, while this trend becomes
ambiguous, or even inverse, in the low-metallicity case.\\
--- Toward a lower metallicity, the extent of the non-LTE departure
tends to decrease, but the departure appears to penetrate deeper
in the atmosphere, which makes the situation rather complex.\\
--- For a very strong damping-dominated case (i.e., lowest $T_{\rm eff}$
and highest metallicity), the departure from LTE shifts toward the upper
atmosphere and the non-LTE effect becomes comparatively insignificant.

\begin{figure}
  \begin{center}
  \hspace{3mm}\psfig{figure=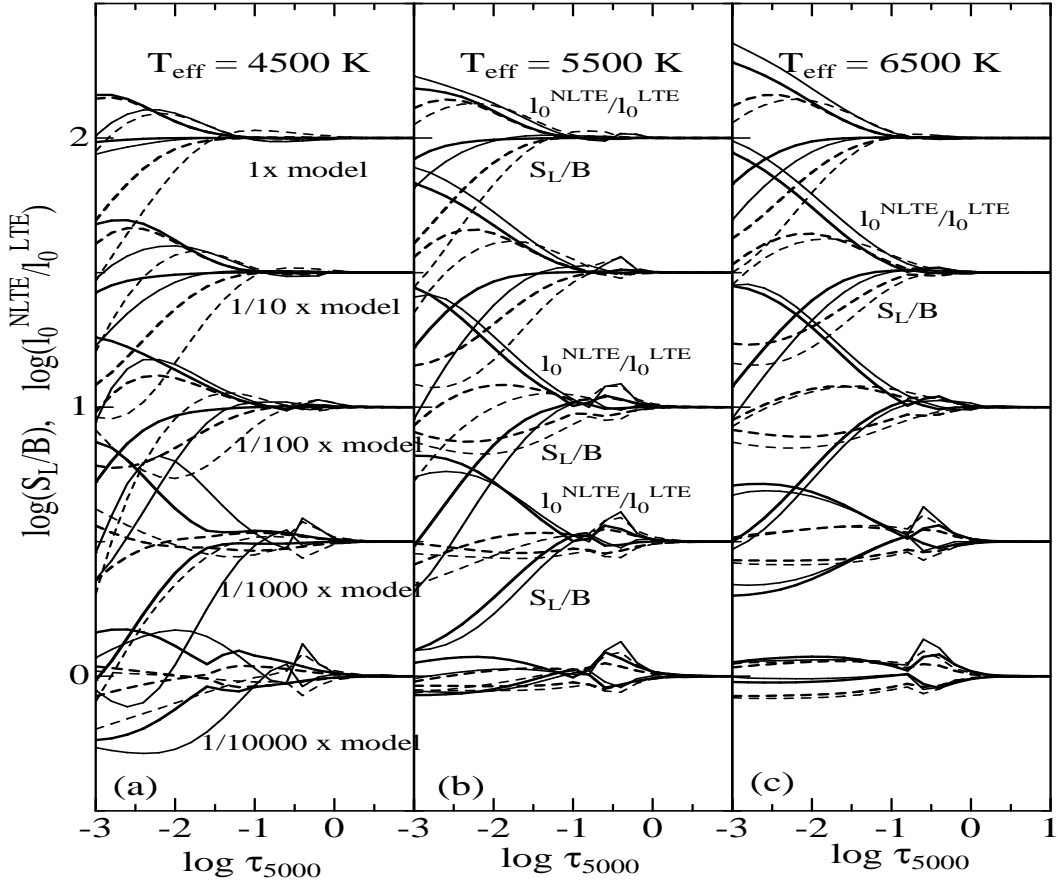,width=155mm,height=155mm,angle=0.0}
\caption{ Ratio of the line source function ($S_{\rm L}$) to the
local Planck function ($B$) and the NLTE-to-LTE line-center
opacity ratio as functions of the standard continuum optical depth
at 5000 $\rm\AA$ computed for models of $T_{\rm eff}$ = 4500~K,
5500~K, and 6500~K. Note that $S_{\rm L}/B$ and $l_{0}^{\rm
NLTE}/l_{0}^{\rm LTE}$ are drawn in the same line-type, but both
are easily discernible since the former (diluted) is generally
lower than the latter (overpopulated). The solid lines show the
results for the 3~$^{2}$S--3p~$^{2}$P$^{\rm o}$ transition of
multiplet 1 (corresponding to Na~{\sc i} 5890/5896 resonance
lines), while those for the 3p~$^{2}$P$^{\rm o}$--3d~$^{2}$D
transition of multiplet 4 (corresponding to Na~{\sc i} 8183/8195
lines) are depicted by dashed lines. In each case, the results for
two different gravity atmospheres are given: The thick lines are
for $\log g = 4$ and the thin lines are for $\log g = 2$,
respectively. Note also that the curves are vertically offset by
an amount of 0.5~dex relative to those of the adjacent metallicity
ones. }
  \end{center}
\end{figure}

\begin{figure}
  \begin{center}
    \hspace{3mm}\psfig{figure=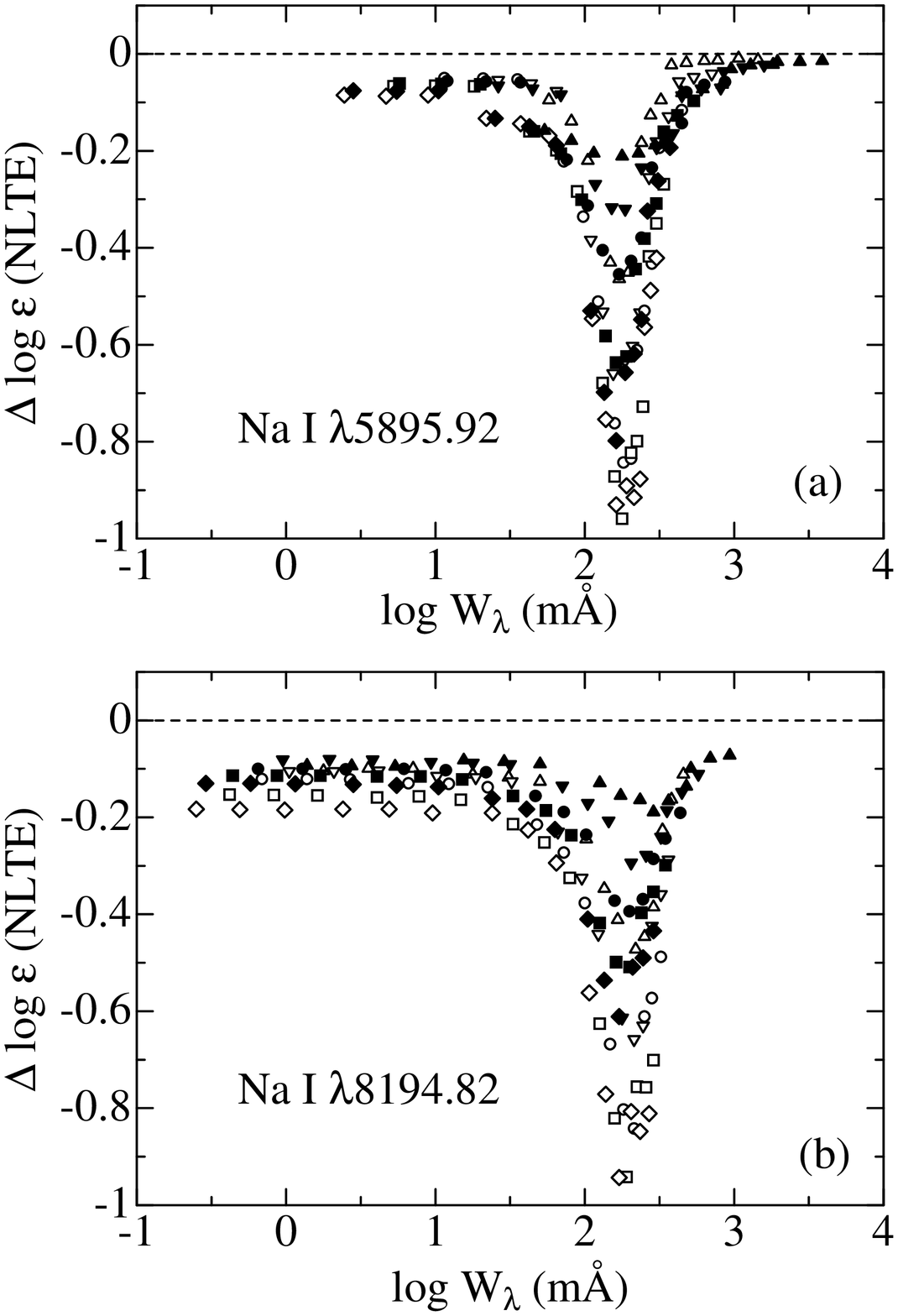,width=155mm,height=155mm,angle=0.0}
\caption{Theoretical non-LTE corrections as functions of the NLTE
equivalent width for representative models, based on the
electronic tables E1 (see subsection 2.3 for details). Different
symbols discern the effective temperature (triangle, inverse
triangle, circle, square, and diamond correspond to $T_{\rm eff}$
= 4500, 5000, 5500, 6000, and 6500~K, respectively) and the
surface gravity (open and closed symbols are for $\log g$ = 2 and
4, respectively) of the adopted model atmospheres. The results
derived for five different model-metallicities ([Fe/H] = 0, $-1$,
$-2$, $-3$, and $-4$), while using the metallicity-scaled sodium
abundance along with perturbations of $\pm 0.3$~dex, are shown
here (though the same symbols are used for different
metallicities/abundances). (a) Na~{\sc i} 5895.92. (b) Na~{\sc i}
8194.82. }
  \end{center}
\end{figure}

\subsection{Grid of Non-LTE Corrections}

Based on the results of these calculations, we computed extensive
grids of the theoretical equivalent-widths and the corresponding non-LTE
corrections for the considered eight lines (Na~{\sc i} 5683, 5688, 5890,
5896, 6154, 6161, 8183, and 8195) for each of the model atmospheres
as follows.

For an assigned sodium abundance ($A^{\rm a}$) and
microturbulence ($\xi^{\rm a}$), we first
calculated the non-LTE equivalent width ($W^{\rm NLTE}$) of the line
by using the computed non-LTE departure coefficients ($b$) for each model
atmosphere. Next, the LTE ($A^{\rm L}$) and NLTE ($A^{\rm N}$) abundances
were computed from this $W^{\rm NLTE}$
while regarding it as if being a given observed equivalent width.
We can then obtain the non-LTE abundance correction, $\Delta$, which is
defined in terms of these two abundances as
$\Delta \equiv A^{\rm N} - A^{\rm L}$.

Strictly speaking, the departure coefficients [$b(\tau)$] for a model 
atmosphere are applicable only to the case of the sodium abundance 
($\log\epsilon_{\rm Na}^{\rm input}$) and the microturbulence (2~km~s$^{-1}$) 
adopted in the non-LTE calculations (cf. subsection 2.1).
Nevertheless, considering the fact that the departure coefficients
(i.e., {\it ratios} of NLTE to LTE number populations) are
(unlike the population itself) not much sensitive to small changes in
atmospheric parameters, we also applied such computed $b$ values to
evaluating $\Delta$ for slightly different $A^{\rm a}$ and $\xi^{\rm a}$
from those fiducial values assumed in the statistical equilibrium
calculations.
Hence, we evaluated $\Delta$ for three $A^{\rm a}$ values
($\log\epsilon_{\rm Na}^{\rm input}$ and $\pm 0.3$ dex perturbation)
as well as three $\xi$ values (2~km~s$^{-1}$ and $\pm 1$~km~s$^{-1}$
perturbation) for a model atmosphere using the same departure coefficients.

We used the WIDTH9 program (Kurucz 1993a), which had been modified
to incorporate the non-LTE departure in the line source function
as well as in the line opacity, for calculating the equivalent
width for a given abundance, or inversely evaluating the abundance
for an assigned equivalent width . The adopted line data ($gf$ values,
radiation damping constants, etc.) are given in table 1.

\begin{table}[]
\caption[]{Adopted atomic data of the Na~{\sc i} lines.}
\begin{center}
\begin{tabular}
{cccccc}\hline \hline
Mult. & Line & $\lambda$ & $\chi$ & $\log gf$ & $\Gamma_{\rm R}$ \\
\hline
1  & $\lambda$5890 & 5889.95 & 0.00 &  +0.12  & 0.62 \\
1  & $\lambda$5896 & 5895.92 & 0.00 & $-0.18$ & 0.62 \\
4  & $\lambda$8183 & 8183.26 & 2.10 &  +0.22  & 1.13 \\
4  & $\lambda$8195 & 8194.82 & 2.10 &  +0.52  & 1.13 \\
5  & $\lambda$6154 & 6154.23 & 2.10 & $-1.56$ & 0.75 \\
5  & $\lambda$6161 & 6160.75 & 2.10 & $-1.26$ & 0.75 \\
6  & $\lambda$5683 & 5682.63 & 2.10 & $-0.67$ & 0.81 \\
6  & $\lambda$5688 & 5688.21 & 2.10 & $-0.37$ & 0.81 \\
\hline
\end{tabular}
\end{center}
Notes. Each of the columns 1--6 give the multiplet number, the
line abbreviation used in this paper, the line wavelength (in
$\rm\AA$), the lower excitation potential (in eV) , the logarithm
of the $gf$ value, and the radiation damping constant in unit of
$10^{8}$~s$^{-1}$. The data presented here are based on table 1 of
Takeda and Takada-Hidai (1994), who consulted the compilation of
Wiese et al. (1969). Each line was treated as if it is a single
line, while neglecting any hyperfine structure (cf. subsection
2.3). See also subsection 2.3 for the treatment of the quadratic
Stark effect damping and of the van der Waals effect damping.
\end{table}

One of the controversial factors in abundance determinations for
late-type stars is the choice of the van der Waals effect damping
constant. Regarding this parameter, we assumed the classical Uns\"old's
(1955) formula unchanged, which means the adoption of
$\Delta\log C_{6} = 0.0$ ($C_{6}$ is connected to the damping width as
$\log\Gamma_{6} = \log\Gamma_{6}^{\rm classical}  + 0.4 \Delta\log C_{6}$).
While this choice is based on our previous empirical investigations
(cf. appendix A of Takeda, Takada-Hidai 1994; subsection 4.2 of Takeda 1995),
we actually confirmed that this choice is reasonable from the analysis of
solar Na~{\sc i} lines (cf. subsection 4.2).
Anyhow, the precise value of this correction is not very essential
unless very strong lines are used, as can be seen from figure 3
where the abundance changes expected by using $\Delta \log C_{6} = 0.5$
are graphically displayed. Namely, the maximum change is
$\sim 0.4 |\Delta \log C_{6}|$ at most occurring only
in the case of very strong damping-dominated lines and
the low-temperature condition where $\Gamma_{6}$ dominates
$\Gamma$ ($\equiv \Gamma_{\rm R} + \Gamma_{4} + \Gamma_{6}$).
Regarding the quadratic Stark effect damping ($\Gamma_{4}$, which is
comparatively insignificant in late-type stars), we followed the
Peytremann's formula (Kurucz 1979, p.8).

\begin{figure}
  \begin{center}
    \hspace{3mm}\psfig{figure=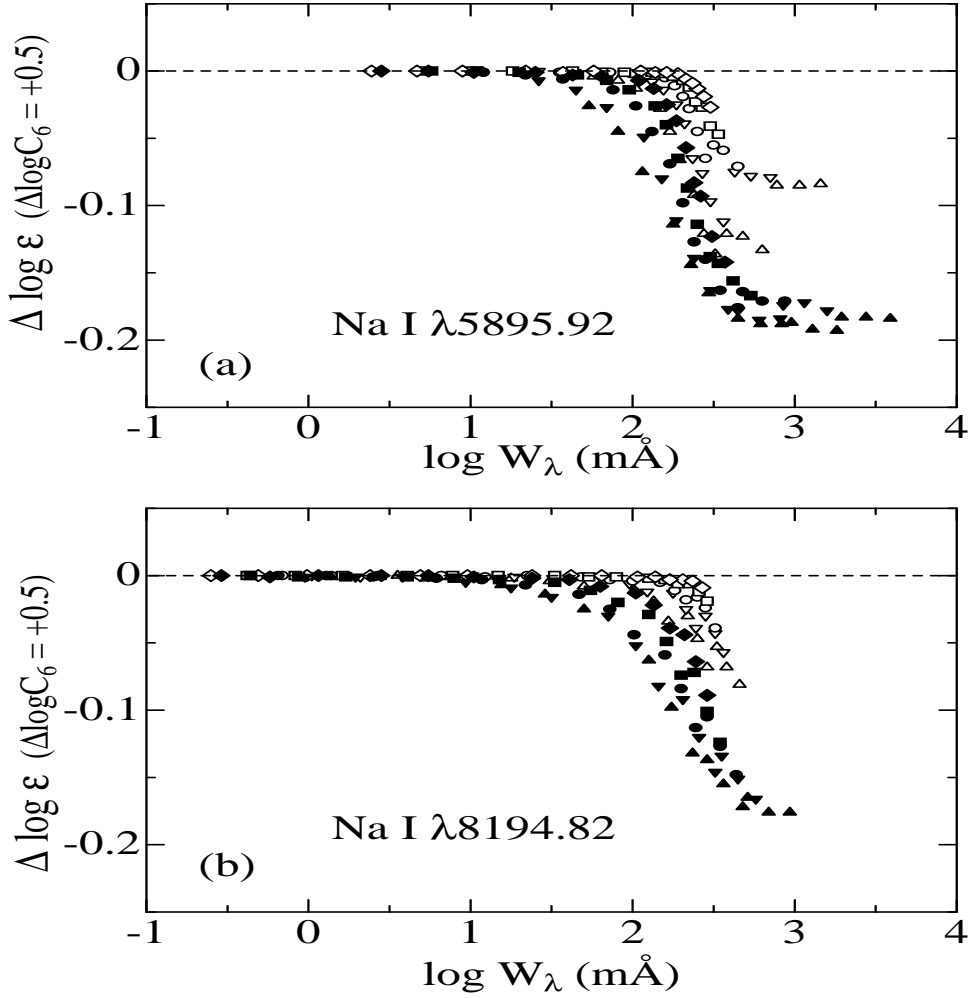,width=155mm,height=155mm,angle=0.0}
\caption{Differences of the resulting non-LTE abundances, when
$\Delta \log C_{6} = +0.5$ (logarithmic correction applied to classical
value of the van der Waals effect damping parameter) was used
instead of the fiducial value of $\Delta \log C_{6} =  0.0$, as
functions of the non-LTE equivalent widths for the representative
cases shown in table 2. See the caption of figure 2 for the
meanings of the symbols. (a) Na~{\sc i} 5895.92. (b) Na~{\sc i}
8194.82. }
   \end{center}
\end{figure}

Note also that we neglected the hyperfine structure (hfs; see, e.g.,
table 4 of McWilliam et al. 1995b for the 5890/5896 lines, and table 1
of Takeda, Takada-Hidai 1994 for the 8195 line) while treating each line
as being purely single, because of a technical reason.\footnote{We decided
to invoke the original WIDTH9 algorithm (applicable to a symmetric single
line) for an equivalent-width calculation instead of using the 
spectral-synthesis technique (which requires a sufficiently fine division 
of a line profile unsuited/unnecessary for very strong lines), since we 
wanted to treat cases of considerably different line strengths in the same 
consistent way (i.e., for constructing the tables of non-LTE corrections 
for different lines over wide parameter ranges).} However, this does not 
make any serious affection (several hundredths dex in most cases of 
actual importance) as illustrated in figure 4.

\begin{figure}
  \begin{center}
    \hspace{3mm}\psfig{figure=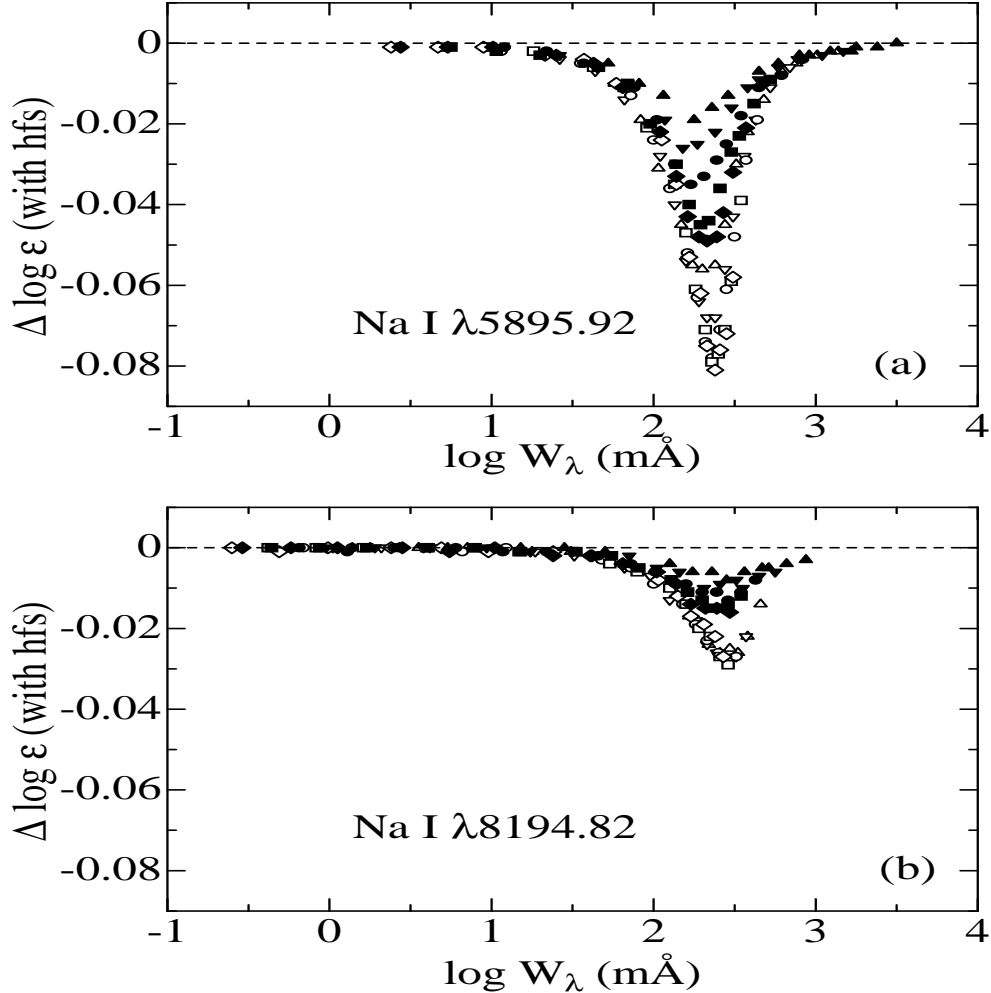,width=155mm,height=155mm,angle=0.0}
\caption{Differences of the resulting non-LTE abundances, when the
hyperfine structure was approximately included instead of the
purely single-line treatment (as basically adopted in this study;
cf. table 1), as functions of the non-LTE equivalent widths for
the representative cases shown in table 2. See the caption of
figure 2 for the meanings of the symbols. (a) Na~{\sc i}
$\lambda$5896 [using the two-component approximation of
$gf(5895.93) = 0.269$ and $gf(5895.95) = 0.376$; cf. McWilliam et
al. 1995b]. (b) Na~{\sc i} $\lambda$8195 [using the two-component
approximation of $gf(8194.79) = 0.332$ and $gf(8194.82) = 2.999$;
cf. Wiese et al. 1969]. }
  \end{center}
\end{figure}

As a demonstrative example of non-LTE corrections, we give
the $\xi$ = 1~km~s$^{-1}$ results for the Na~{\sc i}~5896 and 8195 lines
computed for representative parameters [chosen to be nearly compatible
with Baum{\" u}ller et al.'s (1998) calculations] in table 2,
where we also present the cases of $h = 0, -2, -3$ in addition to the
fiducial $h = -1$ case ($h$ is the logarithmic H~{\sc i} collision correction)
for a comparison. Comparing this table 2 with Baum{\" u}ller et al.'s
(1998) table 2 (where they adopted $10^{h} = 0.05$ or $h = -1.3$),
we can see that our $|\Delta|$ values are quantitatively somewhat
larger than theirs, though the qualitative tendency is in good agreement.
The complete results (for all combinations of $T_{\rm eff}$, $\log g$,
$\xi$ values for each of the 8 lines, though only for the fiducial case of
$h = -1$) are available only in the electronic form. These
tables (named ``tables E1'') are temporarily at the anonymous ftp site of
\begin{verbatim}
ftp://www.ioa.s.u-tokyo.ac.jp/Users/takeda/sodium_paper/
(IP address: 133.11.160.242)
\end{verbatim}
but will be replaced to CDS when the paper is to be published.

\begin{table}[]
\small
\caption{Dependence of the non-LTE effect on the choice of the
correction factor for the H~{\sc i} collision.}
\begin{center}
\begin{tabular}
{ccrcccr@{ }r@{ }r@{ }rr@{ }r@{ }r@{ }r}\hline \hline $T_{\rm
eff}$  & $\log g$ & [Fe/H] & $\xi$ & $A^{\rm a}$ & $\lambda$ &
$W_{0}$ & $W_{-1}$ & $W_{-2}$ & $W_{-3}$ &
$\Delta_{0}$ & $\Delta_{-1}$ & $\Delta_{-2}$ & $\Delta_{-3}$ \\
\hline
5500 & 4.0  & 0.0  & 1.0  & 6.33  & 5895.92 & 602.6  & 616.6  & 616.6  & 631.0  & $-$0.04  & $-$0.06  & $-$0.06  & $-$0.06  \\
5500 & 4.0  & $-$1.0  & 1.0  & 5.33  & 5895.92 & 309.0  & 316.2  & 331.1  & 331.1  & $-$0.10  & $-$0.15  & $-$0.17  & $-$0.18  \\
5500 & 4.0  & $-$2.0  & 1.0  & 4.33  & 5895.92 & 158.5  & 173.8  & 177.8  & 177.8  & $-$0.26  & $-$0.37  & $-$0.40  & $-$0.41  \\
5500 & 4.0  & $-$3.0  & 1.0  & 3.33  & 5895.92 & 85.1  & 89.1  & 91.2  & 89.1  & $-$0.24  & $-$0.33  & $-$0.35  & $-$0.33  \\
5500 & 4.0  & $-$4.0  & 1.0  & 2.33  & 5895.92 & 20.0  & 20.9  & 21.4  & 20.9  & $-$0.03  & $-$0.06  & $-$0.06  & $-$0.05  \\
6500 & 4.0  & 0.0  & 1.0  & 6.33  & 5895.92 & 263.0  & 269.1  & 275.4  & 275.4  & $-$0.18  & $-$0.21  & $-$0.21  & $-$0.22  \\
6500 & 4.0  & $-$1.0  & 1.0  & 5.33  & 5895.92 & 169.8  & 177.8  & 182.0  & 182.0  & $-$0.47  & $-$0.55  & $-$0.56  & $-$0.56  \\
6500 & 4.0  & $-$2.0  & 1.0  & 4.33  & 5895.92 & 109.7  & 114.8  & 114.8  & 114.8  & $-$0.61  & $-$0.71  & $-$0.72  & $-$0.72  \\
6500 & 4.0  & $-$3.0  & 1.0  & 3.33  & 5895.92 & 38.0  & 39.8  & 39.8  & 38.9  & $-$0.13  & $-$0.16  & $-$0.16  & $-$0.16  \\
6500 & 4.0  & $-$4.0  & 1.0  & 2.33  & 5895.92 & 5.1  & 5.4  & 5.4
& 5.4  & $-$0.06  & $-$0.08  & $-$0.08  & $-$0.08  \\ \hline
5500 & 4.0  & 0.0  & 1.0  & 6.33  & 8194.82 & 295.1  & 309.0  & 316.2  & 316.2  & $-$0.15  & $-$0.20  & $-$0.21  & $-$0.22  \\
5500 & 4.0  & $-$1.0  & 1.0  & 5.33  & 8194.82 & 154.9  & 173.8  & 182.0  & 186.2  & $-$0.20  & $-$0.35  & $-$0.42  & $-$0.44  \\
5500 & 4.0  & $-$2.0  & 1.0  & 4.33  & 8194.82 & 57.5  & 66.1  & 69.2  & 70.8  & $-$0.08  & $-$0.20  & $-$0.25  & $-$0.26  \\
5500 & 4.0  & $-$3.0  & 1.0  & 3.33  & 8194.82 & 10.0  & 11.8  & 12.3  & 12.3  & $-$0.03  & $-$0.10  & $-$0.13  & $-$0.13  \\
5500 & 4.0  & $-$4.0  & 1.0  & 2.33  & 8194.82 & 1.1  & 1.3  & 1.4  & 1.4  & $-$0.03  & $-$0.10  & $-$0.12  & $-$0.12  \\
6500 & 4.0  & 0.0  & 1.0  & 6.33  & 8194.82 & 199.5  & 213.8  & 213.8  & 218.8  & $-$0.34  & $-$0.43  & $-$0.47  & $-$0.47  \\
6500 & 4.0  & $-$1.0  & 1.0  & 5.33  & 8194.82 & 107.2  & 117.5  & 123.0  & 123.0  & $-$0.38  & $-$0.55  & $-$0.60  & $-$0.61  \\
6500 & 4.0  & $-$2.0  & 1.0  & 4.33  & 8194.82 & 34.7  & 38.0  & 38.9  & 38.9  & $-$0.12  & $-$0.19  & $-$0.21  & $-$0.21  \\
6500 & 4.0  & $-$3.0  & 1.0  & 3.33  & 8194.82 & 4.9  & 5.5  & 5.6  & 5.6  & $-$0.09  & $-$0.13  & $-$0.14  & $-$0.14  \\
6500 & 4.0  & $-$4.0  & 1.0  & 2.33  & 8194.82 & 0.5  & 0.6  & 0.6  & 0.6  & $-$0.09  & $-$0.13  & $-$0.14  & $-$0.14  \\
\hline
\end{tabular}
\end{center}
Notes. Columns 1--6 are self-explanatory (the units of $T_{\rm
eff}$, $g$, and $\xi$ are K, cm~s$^{-2}$, and km~s$^{-1}$,
respectively). The suffixes ($0$, $-1$, $-2$, $-3$, and $-4$)
appended to $W$ (the non-LTE equivalent width in m$\rm\AA$ calculated 
for the atmospheric parameters and the assigned sodium abundance
given in columns 1--5) and $\Delta$ (the non-LTE abundance
correction) denote the corresponding values of $h$ (the logarithm
of the H~{\sc i} collision correction factor applied to the
classical formula).
\end{table}

\section{Observational Data}
\label{sect:Obs} Our observational data consist of those obtained
at two observatories, Beijing Astronomical Observatory (BAO; 2.2 m
reflector + coude echelle spectrograph at Xinglong station) in P.
R. China and Okayama Astrophysical Observatory (OAO; 1.9 m
reflector + coude HIgh-Dispersion Echelle Spectrograph named
``HIDES'') in Japan.

\subsection{BAO Data}

The BAO spectra ($R \sim 40000$ and S/N $\sim$ 200--400) used here
are such those originally collected for the purpose of Chen et al.'s
(2000) comprehensive analyses on Galactic disk stars. See
subsection 2.2 therein for detailed information on
the observations and the data quality. Chen et al. (2000)
determined the abundance of sodium based on the comparatively weak
Na~{\sc i} 6154/6161 lines, though the used equivalent-widths data 
were not published. For the present study, additional measurements
for five lines (Na~{\sc i} 5683, 5688, 5890, 5896, and 8195; the
8183 line was discarded because of the contamination of telluric
lines) were newly carried out by one of us (Y.-Q. Chen; cf.
subsection 2.3 of Chen et al. 2000 for the measurement method) on
the spectra of 22 F--G stars (+ Moon), which are our BAO program
stars also adopted in our previous potassium analysis (Takeda et
al. 2002)\footnote{One of our present 22 BAO stars, HD~167588, was
not included in Takeda et al.'s (2002) analysis, because there was
a problem in the data quality of the K~{\sc i} 7699 line.}.
Although many lines (except for the 6154/6161 lines) are so strong
in these disk stars and not suitable for sodium abundance
determinations, we tried to derive the abundances from them in
order to see whether the similar behavior of [Na/Fe] may be
reproduced among the different line groups (cf. subsection 5.2).
Yet, it should be kept in mind that difficulties are involved in
measuring the equivalent widths of very strong damping-dominated
lines such as the 5890/5896 D lines, with which errors are more or
less involved. Nevertheless, comparing the solar (Moon) equivalent
width measured by us with those taken from the literature (cf.
table 3), we can see that our measurements are regarded as being
reasonable (i.e., crucial systematic discrepancies are not
observed). The finally obtained BAO equivalent-widths are given in
tables 4 and 5.

\begin{table}[]
\caption{Comparison of the solar equivalent widths (in m$\rm\AA$)
of Na~{\sc i} lines taken from various literature.}
\begin{center}
\begin{tabular}
{crrrrrrrrc}\hline \hline Ref.  & $\lambda$5683 & $\lambda$5688 &
$\lambda$5890 & $\lambda$5896 &
 $\lambda$6154 & $\lambda$6161 & $\lambda$8183 & $\lambda$8195 & Data source\\
\hline
(1) & 100.4 & 131.1 & 837.2 & 545.0 & 38.6 & 59.5 & $\cdots$ & 306.1 & BAO Moon$^{a}$ \\
(2) & 105.3 & 128.6 & $\cdots$ & $\cdots$ & $\cdots$ & $\cdots$ & $\cdots$ & $\cdots$ & B76 solar flux\\
(3) & 106.9 & 127.9 & $\cdots$ & $\cdots$ & $\cdots$ & $\cdots$ & $\cdots$ & $\cdots$ & B76 solar flux\\
(4) & 106.9 & 127.9 & $\cdots$ & $\cdots$ & 36.8 & 58.6 & $\cdots$ & $\cdots$ & B76 solar flux\\
(5) & $\cdots$ & 129.0 & 1125.5 & 736.5 & $\cdots$ & $\cdots$ & 257.3 & 326.6 & solar flux \& disk center$^{b}$\\
(6) & $\cdots$ & $\cdots$ & $\cdots$ & $\cdots$ & 35.0 & 53.7 & $\cdots$ & $\cdots$ & solar disk center$^{c}$\\
(7) & 90.0 & 119.1 & $\cdots$ & $\cdots$ & 40.5 & 58.2 & $\cdots$ & $\cdots$ & K84 solar flux\\
(8) & 121~~ & 144~~ & 830~~ & 640~~ & 41~~ & 63~~ & 254~~ & 328~~ & K84 solar flux\\
(9) & $\cdots$ & $\cdots$ & 756~~ & 597~~ & $\cdots$ & $\cdots$ & 244~~ & 315~~ & K84 solar flux$^{d}$\\
(10) & 103.0 & 128.0 & $\cdots$ & $\cdots$ & 36.0 & $\cdots$ & $\cdots$ & $\cdots$ & From (12)?\\
(11) & $\cdots$ & $\cdots$ & $\cdots$ & $\cdots$ & 38.6 & 59.5 & $\cdots$ & $\cdots$ & K84 solar flux\\
(12) & 103~~ & 128~~ & 765~~ & 570~~ & 36~~ & 53~~ & 239~~ & 322~~ & solar disk center$^{e}$\\
\hline
\end{tabular}
\end{center}
References: (1) This study; (2) Peterson and Carney (1979); (3)
Peterson (1980); (4) Peterson (1981);  (5) Gratton and Sneden
(1987b); (6) Tomkin et al. (1985); (7) Prochaska et al. (2000);
(8) Mashonkina et al. (2000); (9) Takeda (1995); (10) Carretta et
al. (2000);
(11) Sadakane et al. (2002); (12) Holweger (1971).\\ \\
Notes on the data sources:\\
B76 and K84 means the solar flux spectrum atlas
published by Beckers et al. (1976) and Kurucz et al. (1984), respectively.\\
$^{a}$ Moon spectrum observed at Beijing Astronomical Observatory.\\
$^{b}$ Based on two different atlases (see Gratton, Sneden 1987a).\\
$^{c}$ Taken from the values published by Lambert and Luck (1978), which are based on two different atlases.\\
$^{d}$ Inversely computed from the profile fitting solutions.\\
$^{e}$ Based on three different atlases.\\
\end{table}

\begin{table}[]
\scriptsize
\caption{Analysis of Na~{\sc i} 8183/8195 lines based on the BAO
and OAO data.}
\begin{center}
\begin{tabular}{l@{ }l@{ }ccrcr@{ }r@{ }rr@{ }r@{ }rl
}\hline\hline Star & Sp. type & $T_{\rm eff}$ & $\log g$ & [Fe/H]
& $\xi$ & $W_{8183}$ & $A^{\rm N}_{8183}$ & $\Delta_{8183}$ &
$W_{8195}$ & $A^{\rm N}_{8195}$ & $\Delta_{8195}$ & Ref.\\
\hline
\multicolumn{12}{c}{[BAO sample]}\\
Sun & G2 V & 5780 & 4.44  & 0.00  & 1.0  & $\cdots$  & $\cdots$  & $\cdots$  & 306.1  & 6.32  & $-$0.19 & (1) \\
HD~010307 & G1.5 V & 5776 & 4.13  & $-$0.05  & 1.8  & $\cdots$  & $\cdots$  & $\cdots$  & 265.5  & 6.08  & $-$0.31 & (1) \\
HD~019373 & G0 V & 5867 & 4.01  & +0.03  & 1.8  & $\cdots$  & $\cdots$  & $\cdots$  & 274.8  & 6.23  & $-$0.32 & (1)\\
HD~022484 & F9 IV-V & 5915 & 4.03  & $-$0.13  & 2.0  & $\cdots$  & $\cdots$  & $\cdots$  & 242.5  & 5.97  & $-$0.39 & (1) \\
HD~034411 & G1.5 IV-V & 5773 & 4.02  & +0.01  & 1.7  & $\cdots$  & $\cdots$  & $\cdots$  & 275.8  & 6.20  & $-$0.30 & (1) \\
HD~039587 & G0 V & 5805 & 4.29  & $-$0.18  & 2.2  & $\cdots$  & $\cdots$  & $\cdots$  & 273.9  & 5.99  & $-$0.32 & (1) \\
HD~041640 & F5 & 6004 & 4.37  & $-$0.62  & 2.0  & $\cdots$  & $\cdots$  & $\cdots$  & 205.6  & 5.65  & $-$0.39 & (1) \\
HD~049732 & F8 & 6260 & 4.15  & $-$0.70  & 1.9  & $\cdots$  & $\cdots$  & $\cdots$  & 178.3  & 5.63  & $-$0.50 & (1) \\
HD~055575 & G0 V & 5802 & 4.36  & $-$0.36  & 1.6  & $\cdots$  & $\cdots$  & $\cdots$  & $\cdots$  & $\cdots$  & $\cdots$ & (1) \\
HD~060319 & F8 & 5867 & 4.24  & $-$0.85  & 1.6  & $\cdots$  & $\cdots$  & $\cdots$  & 171.4  & 5.41  & $-$0.39 & (1) \\
HD~062301 & F8 V & 5837 & 4.23  & $-$0.67  & 1.7  & $\cdots$  & $\cdots$  & $\cdots$  & 190.1  & 5.53  & $-$0.38 & (1) \\
HD~068146 & F7 V & 6227 & 4.16  & $-$0.09  & 2.1  & $\cdots$  & $\cdots$  & $\cdots$  & 229.4  & 6.01  & $-$0.42 & (1) \\
HD~069897 & F6 V & 6243 & 4.28  & $-$0.28  & 2.0  & $\cdots$  & $\cdots$  & $\cdots$  & 197.5  & 5.77  & $-$0.42 & (1) \\
HD~076349 & G0 & 6004 & 4.21  & $-$0.49  & 2.1  & $\cdots$  & $\cdots$  & $\cdots$  & 210.9  & 5.69  & $-$0.43 & (1) \\
HD~101676 & F6 V & 6102 & 4.09  & $-$0.47  & 2.0  & $\cdots$  & $\cdots$  & $\cdots$  & 214.8  & 5.82  & $-$0.47 & (1) \\
HD~106516 & F5 V & 6135 & 4.34  & $-$0.71  & 1.5  & $\cdots$  & $\cdots$  & $\cdots$  & 168.6  & 5.55  & $-$0.41 & (1) \\
HD~109303 & F8 & 5905 & 4.10  & $-$0.61  & 1.7  & $\cdots$  & $\cdots$  & $\cdots$  & 186.5  & 5.56  & $-$0.43 & (1) \\
HD~118244 & F5 V & 6234 & 4.13  & $-$0.55  & 2.3  & $\cdots$  & $\cdots$  & $\cdots$  & 197.7  & 5.68  & $-$0.50 & (1) \\
HD~142373 & F8 Ve & 5920 & 4.27  & $-$0.39  & 1.5  & $\cdots$  & $\cdots$  & $\cdots$  & $\cdots$  & $\cdots$  & $\cdots$ & (1) \\
HD~142860 & F6 IV & 6227 & 4.18  & $-$0.22  & 2.2  & $\cdots$  & $\cdots$  & $\cdots$  & 211.5  & 5.84  & $-$0.44 & (1) \\
HD~167588 & F8 V & 5894 & 4.13  & $-$0.33  & 1.7  & $\cdots$  & $\cdots$  & $\cdots$  & $\cdots$  & $\cdots$  & $\cdots$ & (1) \\
HD~201891 & F8 V-IV & 5827 & 4.43  & $-$1.04  & 1.6  & $\cdots$  & $\cdots$  & $\cdots$  & 152.6  & 5.21  & $-$0.33 & (1) \\
HD~208906 & F8 V-IV & 5929 & 4.39  & $-$0.73  & 1.5  & $\cdots$  & $\cdots$  & $\cdots$  & 173.9  & 5.47  & $-$0.36 & (1) \\
\hline
\multicolumn{12}{c}{[OAO sample]}\\
HD~6833 & G9 III & 4450 & 1.40  & $-$0.90  & 1.6  & 161.3  & 4.92  & $-$0.41  & 207.9  & 5.10  & $-$0.46 & (2) \\
HD~26297 & G5/G6 Ivw & 4500 & 1.20  & $-$1.60  & 1.7  & 86.4  & 4.24  & $-$0.21  & 115.6  & 4.24  & $-$0.32 & (2) \\
HD~73394 & G5 IIIw & 4500 & 1.10  & $-$1.40  & 1.5  & 127.7  & 4.71  & $-$0.37  & 157.0  & 4.73  & $-$0.46 & (3) \\
HD~76932 & F7/F8 IV/V & 5900 & 4.12  & $-$0.80  & 1.3  & 125.1  & 5.40  & $-$0.38  & 155.6  & 5.37  & $-$0.42 & (4)$^{a}$ \\
HD~88609 & G5 IIIw & 4570 & 0.75  & $-$2.70  & 1.9  & 10.8  & 3.18  & $-$0.10  & 20.1  & 3.19  & $-$0.11 & (4) \\
HD~106516 & F5 V & 6200 & 4.31  & $-$0.70  & 1.1  & 134.8  & 5.66  & $-$0.41  & 167.1  & 5.65  & $-$0.42 & (4)$^{a}$ \\
HD~108317 & G0 & 5300 & 2.90  & $-$2.24  & 1.0  & 26.5  & 3.95  & $-$0.16  & 41.0  & 3.91  & $-$0.18 & (5) \\
HD~122563 & F8 IV & 4590 & 1.17  & $-$2.74  & 2.3  & 15.7  & 3.35  & $-$0.11  & 31.5  & 3.41  & $-$0.12 & (6) \\
HD~140283 & sdF3 & 5690 & 3.69  & $-$2.42  & 0.8  & 6.0  & 3.38  & $-$0.12  & 9.6  & 3.29  & $-$0.13 & (7)$^{a}$ \\
HD~165908 & F7 V & 5900 & 4.09  & $-$0.60  & 1.7  & 143.2  & 5.51  & $-$0.39  & 179.9  & 5.51  & $-$0.43 & (4) \\
HD~167588 & F8 V & 5890 & 4.13  & $-$0.33  & 1.7  & 170.9  & 5.76  & $-$0.38  & 205.3  & 5.72  & $-$0.39 & (1) \\
HD~187111 & G8wvar & 4260 & 0.51  & $-$1.85  & 1.8  & 94.5  & 4.18  & $-$0.19  & 147.6  & 4.41  & $-$0.34 & (6) \\
HD~189322 & G8 III & 4464 & 2.00  & $-$1.57  & 2.0  & 222.1  & 5.37  & $-$0.43  & 260.0  & 5.38  & $-$0.38 & (8)$^{b}$ \\
HD~216143 & G5 & 4525 & 1.00  & $-$2.10  & 2.9  & 50.0  & 3.86  & $-$0.12  & 70.7  & 3.78  & $-$0.15 & (2) \\
HD~221170 & G2 IV & 4425 & 1.00  & $-$2.15  & 1.5  & 53.0  & 3.91  & $-$0.14  & 79.4  & 3.93  & $-$0.20 & (5) \\
BD~+37$^{\circ}$1458 & G0 & 5200 & 3.00  & $-$2.00  & 1.3  & 19.8  & 3.73  & $-$0.15  & 42.7  & 3.86  & $-$0.18 & (2) \\
Procyon & F5 IV-V & 6510 & 3.96  & $-$0.05  & 2.2  & 188.7  & 6.14  & $-$0.51  & 220.3  & 6.08  & $-$0.53 & (9) \\
\hline
\end{tabular}
\end{center}
Notes. Columns 1--6 are self-explanatory as in table 2. $W$ is the
observed equivalent width (in m$\rm\AA$), $A^{\rm N}$ is the
logarithmic non-LTE abundance (in the usual normalization of H =
12), and $\Delta$ is the non-LTE correction ($\equiv A^{\rm N} -
A^{\rm L}$), respectively. Column 13 gives the keys to the
references of the atmospheric parameters (cf. subsection 4.1): (1)
Chen et al. (2000);(2) Fulbright (2000); (3) Luck and Bond
(1985);(4) Takada-Hidai et al. (2002); (5) Pilachowski et al.
(1996); (6) Gratton and Sneden (1994); (7) Nissen et al. (2002);
(8) Alonso et al. (1999); (9) Allende Prieto et al. (2002).\\
$^{a}$ Only the $\xi$ value was taken from (2).\\
$^{b}$ The $\xi$ value was assumed.
\end{table}

\begin{table}[]
\scriptsize
\caption{Analysis of Na~{\sc i} 5683/5688, 5890/5896, and
6154/6160 lines based on the BAO data.}
\begin{center}
\begin{tabular}{lr@{ }r@{ }r@{ }r@{ }r@{ }r@{ }r@{ }r@{ }r@{ }r@{ }r@{ }r@{ }r@{ }r@{ }r@{ }r@{ }r@{ }r
}\hline\hline Star & $W_{5683}$ & $A^{\rm N}_{5683}$ &
$\Delta_{5683}$ & $W_{5688}$ & $A^{\rm N}_{5688}$ &
$\Delta_{5688}$ & $W_{5890}$ & $A^{\rm N}_{5890}$ &
$\Delta_{5890}$ & $W_{5896}$ & $A^{\rm N}_{5896}$ &
$\Delta_{5896}$ & $W_{6154}$ & $A^{\rm N}_{6154}$ &
$\Delta_{6154}$ &
$W_{6161}$ & $A^{\rm N}_{6161}$ & $\Delta_{6161}$ \\
\hline
\multicolumn{19}{c}{[BAO sample]}\\
Sun & 100.4  & 6.28  & $-$0.11  & 131.1 & 6.33  & $-$0.13  & 837.2  & 6.41  & $-$0.05  & 545.0  & 6.27  & $-$0.07  & 38.6  & 6.31  & $-$0.04  & 59.5  & 6.32  & $-$0.06 \\
HD~010307 & $\cdots$  & $\cdots$  & $\cdots$  & 134.3 & 6.26  & $-$0.16  & 779.2  & 6.43  & $-$0.07  & 506.5  & 6.30  & $-$0.09  & 39.4  & 6.29  & $-$0.05  & 60.6  & 6.28  & $-$0.07 \\
HD~019373 & $\cdots$  & $\cdots$  & $\cdots$  & 139.6 & 6.39  & $-$0.19  & $\cdots$  & $\cdots$  & $\cdots$  & $\cdots$  & $\cdots$  & $\cdots$  & 45.8  & 6.44  & $-$0.05  & 65.3  & 6.40  & $-$0.07 \\
HD~022484 & $\cdots$  & $\cdots$  & $\cdots$  & 116.9 & 6.12  & $-$0.17  & 458.5  & 6.02  & $-$0.12  & 402.6  & 6.16  & $-$0.13  & 32.1  & 6.23  & $-$0.06  & 47.1  & 6.17  & $-$0.07 \\
HD~034411 & $\cdots$  & $\cdots$  & $\cdots$  & 133.6 & 6.30  & $-$0.17  & 795.3  & 6.52  & $-$0.07  & 505.6  & 6.37  & $-$0.08  & 39.7  & 6.30  & $-$0.05  & 65.6  & 6.36  & $-$0.07 \\
HD~039587 & $\cdots$  & $\cdots$  & $\cdots$  & 118.1 & 6.01  & $-$0.13  & 625.6  & 6.11  & $-$0.09  & 459.5  & 6.06  & $-$0.12  & 24.0  & 6.01  & $-$0.05  & 48.4  & 6.11  & $-$0.06 \\
HD~041640 & $\cdots$  & $\cdots$  & $\cdots$  & 77.2 & 5.69  & $-$0.12  & 413.1  & 5.69  & $-$0.16  & 316.3  & 5.64  & $-$0.21  & 9.1  & 5.63  & $-$0.07  & 21.8  & 5.75  & $-$0.07 \\
HD~049732 & $\cdots$  & $\cdots$  & $\cdots$  & 73.7 & 5.78  & $-$0.15  & 343.8  & 5.80  & $-$0.24  & 260.8  & 5.65  & $-$0.34  & 7.7  & 5.65  & $-$0.08  & 17.3  & 5.74  & $-$0.09 \\
HD~055575 & $\cdots$  & $\cdots$  & $\cdots$  & 93.2 & 5.82  & $-$0.12  & $\cdots$  & $\cdots$  & $\cdots$  & 412.6  & 5.90  & $-$0.11  & 21.6  & 5.96  & $-$0.05  & 36.2  & 5.95  & $-$0.06 \\
HD~060319 & $\cdots$  & $\cdots$  & $\cdots$  & 59.7 & 5.45  & $-$0.11  & 354.5  & 5.42  & $-$0.18  & 270.9  & 5.36  & $-$0.23  & 6.3  & 5.40  & $-$0.07  & 17.8  & 5.60  & $-$0.08 \\
HD~062301 & $\cdots$  & $\cdots$  & $\cdots$  & 71.8 & 5.58  & $-$0.12  & 407.8  & 5.60  & $-$0.14  & 285.8  & 5.43  & $-$0.21  & 10.2  & 5.61  & $-$0.07  & 19.9  & 5.64  & $-$0.07 \\
HD~068146 & $\cdots$  & $\cdots$  & $\cdots$  & 115.9 & 6.24  & $-$0.17  & 436.3  & 6.19  & $-$0.15  & 352.0  & 6.20  & $-$0.18  & 24.6  & 6.23  & $-$0.06  & 43.3  & 6.25  & $-$0.07 \\
HD~069897 & $\cdots$  & $\cdots$  & $\cdots$  & 94.1 & 6.01  & $-$0.14  & 374.0  & 5.92  & $-$0.18  & 306.3  & 5.92  & $-$0.23  & 14.5  & 5.96  & $-$0.06  & 29.1  & 6.02  & $-$0.07 \\
HD~076349 & $\cdots$  & $\cdots$  & $\cdots$  & 87.1 & 5.79  & $-$0.14  & 401.0  & 5.74  & $-$0.17  & 318.0  & 5.71  & $-$0.22  & 14.7  & 5.85  & $-$0.07  & 29.8  & 5.92  & $-$0.08 \\
HD~101676 & $\cdots$  & $\cdots$  & $\cdots$  & 93.0 & 5.93  & $-$0.16  & $\cdots$  & $\cdots$  & $\cdots$  & 294.5  & 5.76  & $-$0.27  & 15.3  & 5.92  & $-$0.07  & 31.2  & 5.99  & $-$0.08 \\
HD~106516 & $\cdots$  & $\cdots$  & $\cdots$  & 69.9 & 5.71  & $-$0.13  & 313.5  & 5.52  & $-$0.21  & 260.6  & 5.56  & $-$0.26  & 9.0  & 5.68  & $-$0.07  & 22.1  & 5.82  & $-$0.08 \\
HD~109303 & $\cdots$  & $\cdots$  & $\cdots$  & 78.6 & 5.70  & $-$0.14  & 382.4  & 5.66  & $-$0.17  & 279.4  & 5.53  & $-$0.23  & 8.1  & 5.53  & $-$0.07  & 32.2  & 5.93  & $-$0.08 \\
HD~118244 & $\cdots$  & $\cdots$  & $\cdots$  & 91.9 & 5.93  & $-$0.16  & 353.7  & 5.78  & $-$0.25  & 281.9  & 5.68  & $-$0.34  & 9.7  & 5.75  & $-$0.08  & 29.5  & 6.00  & $-$0.09 \\
HD~142373 & 68.1  & 5.89  & $-$0.10  & $\cdots$ & $\cdots$  & $\cdots$  & 407.0  & 5.75  & $-$0.12  & 401.1  & 6.03  & $-$0.13  & 12.6  & 5.75  & $-$0.06  & 26.5  & 5.84  & $-$0.07 \\
HD~142860 & $\cdots$  & $\cdots$  & $\cdots$  & 106.3 & 6.11  & $-$0.16  & 449.6  & 6.18  & $-$0.16  & 317.6  & 5.98  & $-$0.24  & 17.6  & 6.05  & $-$0.06  & 44.0  & 6.25  & $-$0.08 \\
HD~167588 & 75.1  & 5.96  & $-$0.11  & 97.4 & 5.92  & $-$0.15  & 423.4  & 5.84  & $-$0.13  & 368.4  & 5.96  & $-$0.14  & 19.1  & 5.95  & $-$0.06  & 37.6  & 6.02  & $-$0.07 \\
HD~201891 & $\cdots$  & $\cdots$  & $\cdots$  & 43.3 & 5.20  & $-$0.09  & 328.1  & 5.14  & $-$0.19  & 209.3  & 4.82  & $-$0.28  & 7.4  & 5.46  & $-$0.07  & 9.6  & 5.28  & $-$0.07 \\
HD~208906 & 52.1  & 5.68  & $-$0.09  & $\cdots$ & $\cdots$  & $\cdots$  & $\cdots$  & $\cdots$  & $\cdots$  & $\cdots$  & $\cdots$  & $\cdots$  & 9.2  & 5.61  & $-$0.07  & 21.3  & 5.72  & $-$0.07 \\
\hline
\end{tabular}
\end{center}
See the notes to table 4 for the meanings of the $W$, $A^{\rm N}$,
and $\Delta$.
\end{table}

\subsection{OAO Data}

The OAO data are based on the near-IR spectra of 17 mostly metal-poor
disk/halo stars (dwarfs and giants being mixed), which have been collected
by using HIDES during our observing runs in 2001--2002.\footnote{Actually,
the primary motivation of these observations was to study the
abundance of sulfur based on the near-IR S~{\sc i} lines.
Nevertheless, we could apply those spectra to the present purpose,
since they cover the wavelength region of the Na~{\sc i} 8183/8195 lines.}
The slit width of 250~$\mu$m ($0.''95$) was set to yield
a spectral resolution of $R \sim 50000$.
By using the single 4K$\times$2K CCD (pixel size of
13.5~$\mu$m $\times$ 13.5~$\mu$m),
a wavelength span of $\sim 1100 \; \rm\AA$ could be covered
at one time. Most of our observations were made in the wavelength
region of $\sim$ 7700--8800 $\rm\AA$.

The data reduction was performed with the IRAF\footnote{IRAF is
distributed by the National Optical Astronomy Observatories,
which is operated by the Association of Universities for  Research
in Astronomy, Inc., under cooperative agreement with the National
Science Foundation.} {\tt echelle} package, following the standard
procedure for extracting one-dimensional spectra.
We also removed the telluric lines using the {\tt telluric} task
of IRAF and the spectra of a rapid rotator.
The S/N ratios of the resulting spectra, differing from
star to star, are typically of the order of 200--300.
The resulting OAO equivalent widths of the Na~{\sc i} 8183/8195
lines based on these spectra, which were measured by the Gaussian
fitting or direct integration method using the {\tt splot} task
of IRAF, are presented in table 4.
Note that, although two stars (HD~106516 and HD~167588) are common
to BAO and OAO samples, they were treated independently.

\section{Abundance Analyses}
\label{sect:Abundance}

\subsection{Atmospheric Parameters}

First we need to establish the four atmospheric parameters
[$T_{\rm eff}$ (effective temperature), $\log g$ (surface gravity),
[Fe/H] (metallicity; represented by the Fe abundance relative to the Sun),
and $\xi$ (microturbulent velocity dispersion)], which are necessary for
constructing the model atmosphere and deriving the abundance
from an equivalent width.
Regarding the 22 BAO stars, the parameter data determined by
Chen et al. (2000) were used unchanged, while the reasonably known
solar parameters of (5780~K, 4.44, 0.0, 1.0 km~s$^{-1}$) were applied
to the analysis of the BAO Moon data.
Meanwhile, the atmospheric parameters of 17 OAO stars were taken
from the published values found in various literature (cf. column 13
of table 4); when two or more references were available for the same star,
we selected an appropriate value according to our personal judgment.
The finally adopted values are given in table 4.

As for the model atmospheres, Kurucz's (1993a) grid of ATLAS9 models
was used as in the case of non-LTE calculations,
based on which the model of each star was obtained by a three-dimensional
interpolation with respect to $T_{\rm eff}$, $\log g$, and [Fe/H].
Similarly, the depth-dependent departure coefficients ($b$) of
Na~{\sc i} levels computed for the grid of models
(cf. section 2) were interpolated in terms of $T_{\rm eff}$, $\log g$,
and [Fe/H], in order to evaluate the $S_{\rm L}(\tau)/B(\tau)$ and
$l_{0}^{\rm NLTE}(\tau)/l_{0}^{\rm LTE}(\tau)$ ratios for each star.

\subsection{Abundance Determination}

The procedures for determining the sodium abundances from
the equivalent widths of Na~{\sc i} lines are the same as already
explained in subsection 2.3 (i.e., modified WIDTH9 program, line data
given in table 1, $\Delta\log C_{6} = 0$, neglecting the hfs effect).

The resulting non-LTE abundance ($A^{\rm N}$) and
the non-LTE abundance correction ($\equiv A^{\rm N} - A^{\rm L}$)
derived for each line and for each star are presented in tables 4
(8183/8195 lines) and 5 (5683/5688, 5890/5896, and 6154/6161 lines).
As can be seen from these two tables, the solar non-LTE abundances derived
from the BAO Moon equivalent widths of Na~{\sc i} 8195 (6.32),
5683 (6.28), 5688 (6.33), 5890 (6.41), 6154 (6.31), and 6160 (6.32)
are in remarkable agreement with each other; moreover,
their average of $6.32 (\sigma = 0.04)$ is essentially the same as
the standard solar sodium abundance of 6.33 (Anders, Grevesse 1989).
Accordingly, we confirmed the internal consistency of using
$\Delta\log C_{6} = 0.0$, as far as our present analysis
using ATLAS9 model atmospheres is concerned. [Actually,
empirical determination of this correction significantly depends
on the choice of the atmospheric models, as demonstrated in
subsection 4.2 of Takeda (1995).]

The abundance changes caused by uncertainties in the atmospheric
parameters are quantitatively similar to the case of potassium
described in section 5 of Takeda et al. (2002), since K
and Na have quite similar atomic structures to each other
(alkali atom with one valence electron; ionization potential
of 4--5 eV) and most atoms are in the once-ionized stage.
Hence, even for changes of $\Delta T_{\rm eff} \sim \pm 200$~K or
$\Delta \log g \sim \pm 0.3$~dex (we expect that internal errors
in the parameters of our program stars, especially for BAO sample
stars, are smaller than these), the extents of the resulting
abundance variations are $\la 0.2$~dex, as can be seen from
table 1 of Takeda et al. (2002). The effect of changing the
microturbulence, which influences only the lines at the flat part
of the curve of growth ($W_{\lambda} \sim$  100--300 m$\rm\AA$;
cf. figure 4) can also be assessed by inspecting that table.

The [Na/Fe] ratios (Na-to-Fe logarithmic abundance ratio relative to
the Sun) can be obtained based on the results in tables 4 and 5 as
$A^{\rm N}({\rm star}) - 6.33 - [{\rm Fe}/{\rm H}]$,
where we assumed Anders and Grevesse's (1989) value of 6.33
as the reference solar Na abundance.
The resulting [Na/Fe] vs. [Fe/H] relations derived from each of the
four multiplets, along with the corresponding non-LTE corrections,
are displayed in figure 5.

\begin{figure}
  \begin{center}
    \hspace{3mm}\psfig{figure=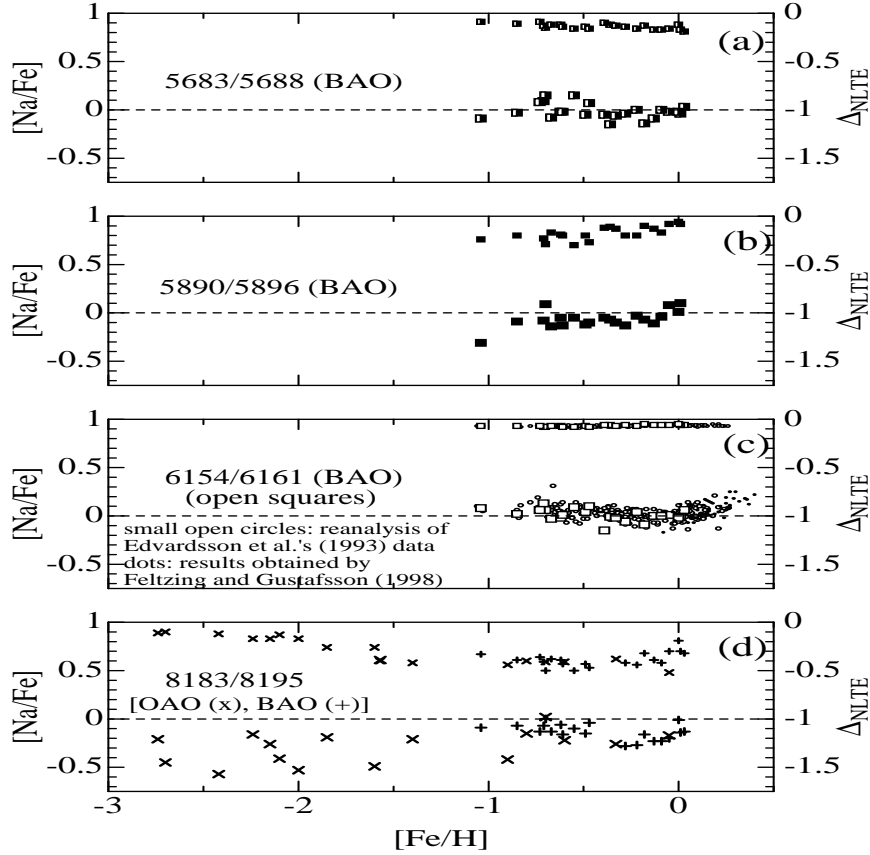,width=120mm,height=120mm,angle=0.0}
\caption{The [Na/Fe] values resulting from the non-LTE analysis of
our BAO and OAO data (larger symbols; scale in the left-hand axis)
and the corresponding non-LTE corrections (smaller symbols; scale
in the right-hand axis), plotted as functions of [Fe/H]; these are
based on the data presented in tables 4 and 5. (a)
$\lambda$5683/$\lambda$5688 lines of multiplet 6, plotted in
half-filled squares (BAO). (b) $\lambda$5890/$\lambda$5896 lines
of multiplet 1, plotted in filled squares (BAO). (c)
$\lambda$6154/$\lambda$6161 lines of multiplet 5, plotted in open
squares (BAO). The results of our reanalysis of Edvardsson et
al.'s (1993) data (also depicted in figure 7a) and the results of
Feltzing and Gustafsson (1998) (the LTE [Na/Fe] values obtained by
themselves; their equivalent-width data are not published) are
also shown with small open circles and dots, respectively, for a
comparison. (d) $\lambda$8183/$\lambda$8195 lines of multiplet 4,
plotted in Greek crosses (BAO) and St. Andrew's crosses (OAO),
respectively. }
  \end{center}
\end{figure}

Note that when two lines of the same multiplet are available,
we averaged the abundances ([Na/Fe]) as well as the
non-LTE corrections for both lines and such derived mean results for
the multiplet are shown.

\subsection{Literature Data Analysis}

We also tried to reanalyze the published equivalent-width data
(for the Na~{\sc i} 5683/5688, 5890/5896, 6154/6161, and 8183/8195
lines) of late-type stars in the Galactic disk as well as in the halo
covering a wide range of metallicities, while taking account of
the non-LTE effect.
In searching for the papers including the observational data we need,
the extensive references quoted by Timmes et al. (1995), Samland (1998),
and Goswami and Prantzos (2000) were quite helpful. Although
our literature survey is not complete, we consider that we have
picked up most of the important works done after 1980's.

These data were analyzed just in the same way as we did for our
BAO/OAO data (cf. subsections 4.1 and 4.2).
The atmospheric parameters were taken from the same paper
as that presenting the equivalent-width data as far as possible.\footnote{
The exceptional cases are as follows:
Gratton and Sneden's (1987b) EW data were analyzed with the parameters
taken from Gratton and Sneden (1987a).
Peterson's (1989) EW data were analyzed with the parameters
of Peterson et al. (1990).
Zhao and Magain's (1990b) EW data were analyzed with ($T_{\rm eff}$,
$\log g$, and $\xi$) values taken from Magain (1989) and [Fe/H] values
taken from Zhao and Magain(1990a).
McWilliam et al.'s (1995a) EW data were analyzed with the parameters
taken from McWilliam et al. (1995b).
} By inspecting those original papers,
we found that abundance analyses done before 1990 tend to be based on
rather coarsely determined parameters (e.g., using rounded $T_{\rm eff}$
or $\log g$ values; assuming the same $\xi$ values for all program stars, etc.)
compared to the more recent studies carried out this decade.
Accordingly, we decided to present the results of the reanalyzes
on the data before and after 1990 separately.
Figure 6 shows the resulting [Na/Fe] vs. [Fe/H] relations
and the corresponding non-LTE corrections derived from the older data
before 1990, while those obtained from the analyses of the recent data
after 1990 are displayed in figure 7.\footnote{Note that the BAO results 
are shown only for the 6156/6161 lines in figure 7, because those for 
the other strong lines are comparatively less reliable.}

The same results as those in figure 7a (the reanalysis results based on 
the new data after 1990) are also depicted in figure 8a, where the [Na/Fe] 
data are plotted with a more compressed vertical scale (for the purpose 
of clarifying the global tendency) and low- and high-gravity stars 
are distinguished by different symbols. In addition, figure 8b shows 
the $T_{\rm eff}$ vs. $\log g$ plots relevant to these data; it can be seen 
from this figure that most of the high-gravity stars are comparatively 
higher $T_{\rm eff}$ stars of mid-F through late-G type, while low-gravity 
stars mainly consist of late-G through early-K giants of lower $T_{\rm eff}$ 
and show a positive correlation between $T_{\rm eff}$ and $\log g$ 
(reflecting the evolutionary sequence).
We may state from an inspection of figure 8a that any meaningful 
systematic $\log g$-dependent difference does not exist in the 
[Na/Fe] values; however, the biased sample (i.e., very low-metallicity 
stars tend to be low-gravity giants, and vice versa) prevents us from 
making any definite argument regarding this point.

\begin{figure}
  \begin{center}
    \hspace{3mm}\psfig{figure=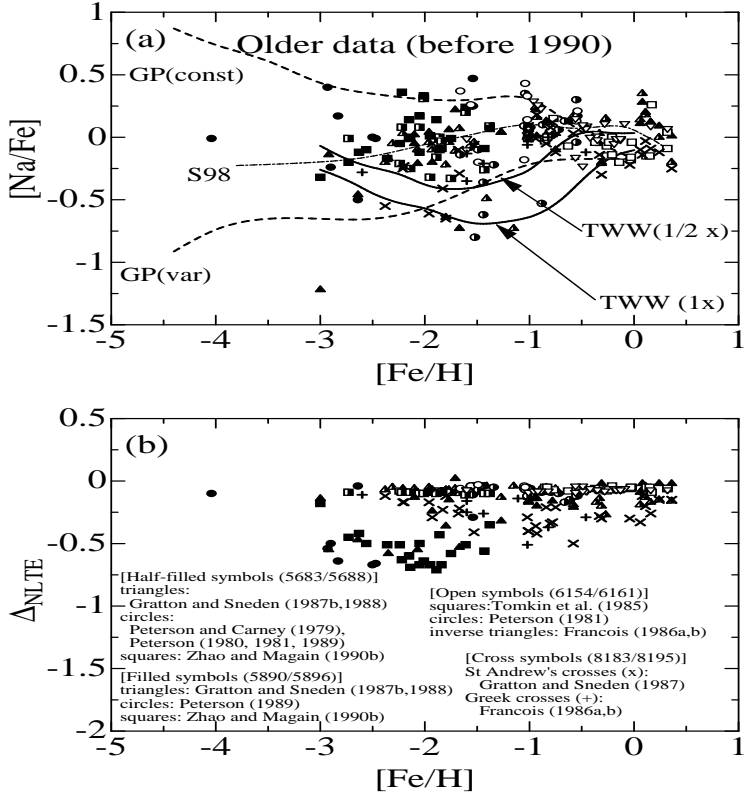,width=120mm,height=120mm,angle=0.0}
\caption{(a) [Na/Fe] vs. [Fe/H] relation resulting from the non-LTE
reanalysis of the rather old data published before 1990. See panel
(b) for the data sources of equivalent widths and the
corresponding symbols. Lines show the representative theoretical
predictions: Dash-dotted line (S98)--- taken from figure 8 of
Samland (1998), dashed line (TWW) --- taken from figure 17 of
Timmes et al. (1995) corresponding to the two cases for the
standard Woosley and Weaver's (1995) Fe yield from massive stars
and the reduced one by a factor of 2 (which may be more reasonable
according to their suggestion), dashed line --- taken from figure
7 of Goswami and Prantzos (2000), for the two cases of Na yield;
time-independent constant yield [GP(const); only for a comparison 
purpose] and the realistic metallicity-dependent Na yield [GP(var)]. 
(b) The corresponding NLTE corrections used for deriving the [Na/Fe] 
values shown in panel (a), plotted as functions of [Fe/H]. }
  \end{center}
\end{figure}

\begin{figure}
  \begin{center}
    \hspace{3mm}\psfig{figure=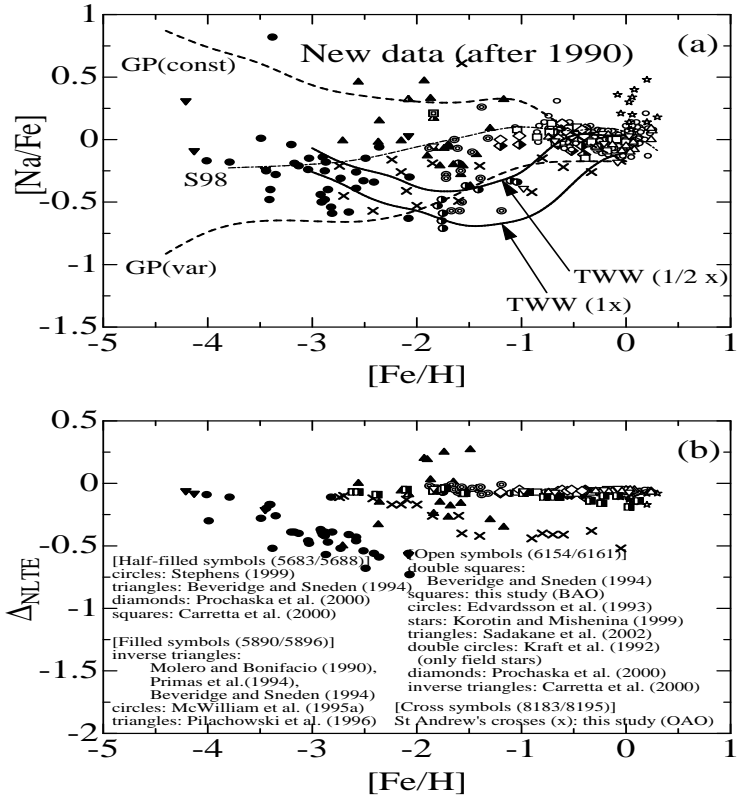,width=120mm,height=120mm,angle=0.0}
\caption{(a) [Na/Fe] vs. [Fe/H] relation resulting from the
non-LTE reanalysis based on the recently published new data after
1990. The data point for CS 22949-037 ([Na/Fe] = +1.94 at
[Fe/H] = $-3.99$) derived from the reanalysis of McWilliam et al.'s
(1995a) EW data is outside the plot range of this figure 
(cf. figure 8a for a more global view including this anomalous point).
(b) The corresponding NLTE corrections used for deriving the
[Na/Fe] values shown in panel (a), plotted as functions of [Fe/H].
Otherwise, the same as in figure 6; see the caption therein for
more details. The results of the reanalysis of Pilachowski et
al.'s (1996) 5890/5896 data for the low-gravity giants/supergiants
(filled triangles) may suffer rather large uncertainties and
should not be seriously taken (cf. the footnote in subsection 5.3).
}
  \end{center}
\end{figure}

\begin{figure}
  \begin{center}
    \hspace{3mm}\psfig{figure=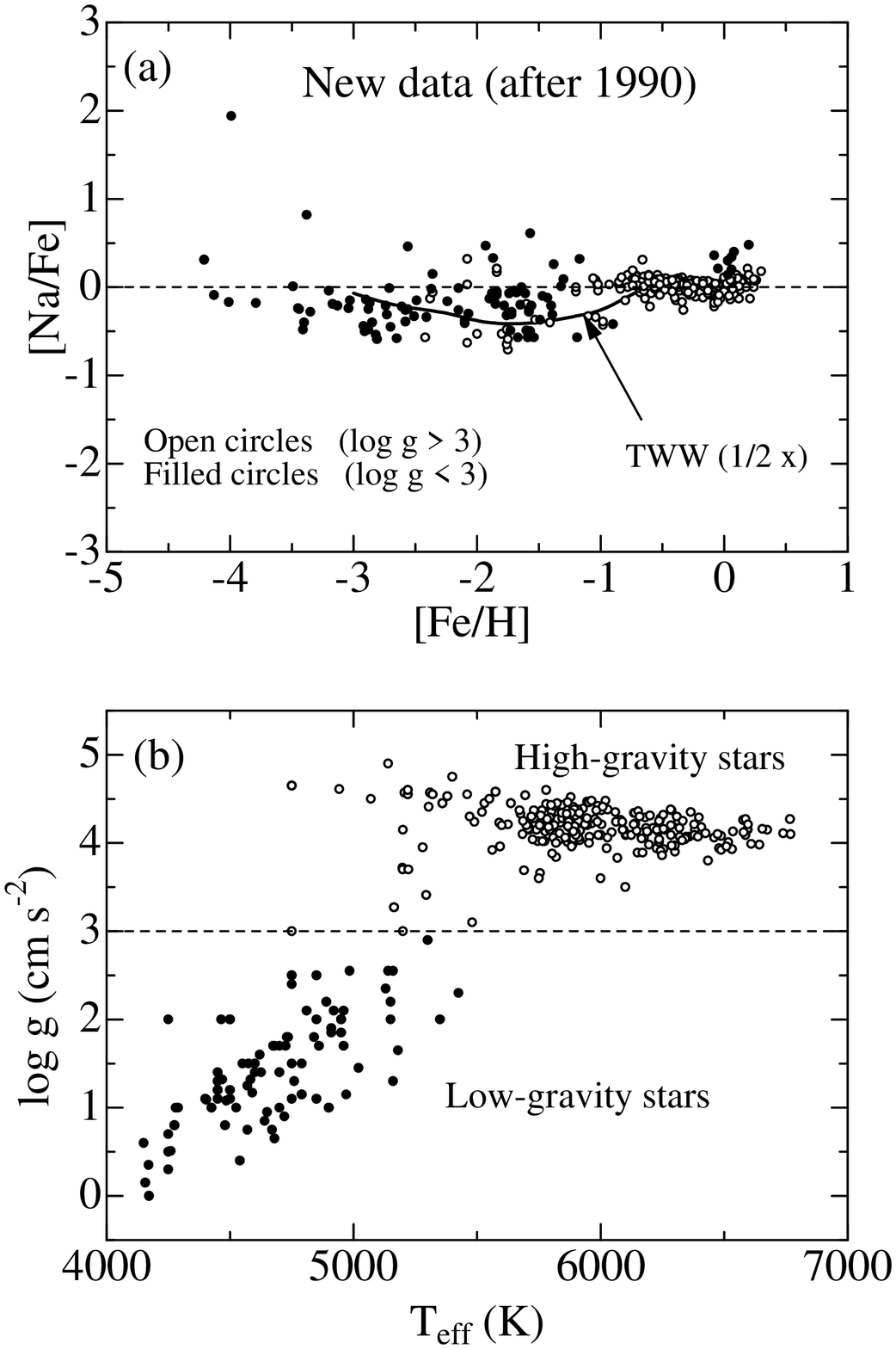,width=120mm,height=120mm,angle=0.0}
\caption{(a) [Na/Fe] vs. [Fe/H] relation resulting from the
non-LTE reanalysis of the new literature data after 1990.
The plotted data are essentially the same as in figure 7a,
but dwarfs and giants are discriminated by different symbols:
open circles --- high-gravity stars ($\log g \ge 3$),
filled clrcles --- low-gravity stars ($\log g < 3$).
The theoretical prediction of Timmes et al. (1995) (the case of
reduced Fe yield denoted by ``TWW (1/2 $\times$)''; cf. the caption
of figure 6a), which appears to match the observed trend most 
satisfactorily, is also drawn for a comparison.
(b) $T_{\rm eff}$ vs. $\log g$ relation for the sample stars
relevant to the data plots in (a). The meanings of the symbols
as the same as in (a).
}
  \end{center}
\end{figure}

The details of these analyses (the data of the used equivalent
widths and the adopted parameter values, the resulting
non-LTE abundances or [Na/Fe] values with the non-LTE corrections,
given for each line/multiplet and for each star) are available
only electronically as ``tables E2''(from the ftp site mentioned
at the end of subsection 2.3).

\section{Discussion}
\label{sect:Disc}

\subsection{Non-LTE Effect}

Figures 5, 6, and 7 clearly show how the non-LTE effect
acts on different Na~{\sc i} lines in sodium abundance
determinations for stars of various metallicities.

Roughly speaking, a simple principle appears to hold:
The non-LTE effect (negative corrections to the corresponding
LTE abundances) tends to be most important for rather strong
saturated lines (with equivalent widths of 100--300 m$\rm\AA$)
on the flat part of the curve of growth, and it becomes
progressively insignificant towards decreasing strengths
(i.e., unsaturated weak lines on the linear part) as well as
increasing strengths (i.e., extremely strong damping-dominated
lines in the damping part), reflecting the importance of the
core-deepening effect caused by the dilution of $S_{\rm L}$
as already mentioned in subsection 2.2 (cf. figure 2).

This naturally explains the behaviors of the extent of
the non-LTE corrections for each line at different
metallicities; for example, the reason why the weak 6154/6161
lines in disk stars show almost negligible non-LTE effects
(cf. figure 5c), why the $|\Delta|$ values for the 8183/8195
attain maximum at [Fe/H] $\sim -1$ (cf. figures 5d, 6b, and 7b),
and why the $|\Delta|$ values for the 5890/5896 lines
systematically decreases from [Fe/H] $\sim -2$ to $\sim -4$
(cf. figure 7b).

We should note, however, that the situation more or less
differs from line to line, reflecting their individual
properties. For example, lines at longer wavelengths tend
to suffer larger non-LTE effects because of the decreasing
sensitivity of the Planck function to the temperature
(i.e., the minimum allowed LTE residual intensity tends to
be raised and thus the difference between LTE and non-LTE
becomes more significant). Hence, we can understand the fact
that the near-IR 8183/8195 lines show relatively larger
non-LTE effects compared to the other lines at similar
line-strengths.

Based on figures 5--7, we can assess the importance of
the non-LTE effect on different lines in sodium
abundance determinations as follows:
The 6154/6161 lines are the best abundance indicators,
for which the non-LTE effect is practically insignificant
(i.e., $\la 0.1$ dex) and may be neglected, though they
are usable/visible only at $-2 \la$ [Fe/H] $\la 0.5$).
The 5683/5688 lines may also be tolerably analyzed with LTE
at the metallicity of $-3 \la$ [Fe/H] $\la -1$.
Regarding the 5890/5896 and 8183/8195 lines, which are
important for diagnosing the Na abundance behaviors of
very metal-deficient stars because of their strengths
(especially, the 5890/5896 D lines are the only tool for
investigating the extremely metal-poor regime of
$-4 \la$ [Fe/H] $\la -3$), we consider that properly
taking account of the non-LTE effect is necessary.

\subsection{Disk Stars}

The overall behavior of the Na abundances for disk stars
($-1 \la$ [Fe/H] $\la +0.4$) has long been known
to nearly scale with Fe; i.e., [Na/Fe] $\sim 0$ (e.g.,
Wallerstein 1962; Tomkin et al. 1985). There remains
almost no doubt for this broad description, which may be
confirmed also in our figures 5--7.\footnote{Note that figures
6--8 include low-gravity giants of population I [e.g.,
reanalysis results of Korochin and Mishenina's (1999) data
shown by open stars in figure 7] showing markedly positive
[Na/Fe] values in contrast to others. These stars are expected
to have suffered evolution-induced Na enrichment in the envelope
caused by the dredge-up of NeNa cycle products in the H-burning
shell, such as the case of supergiants (Takeda, Takada-Hidai 1994).
Hence, their abundance characteristics should be discriminated
from those related to the Galactic chemical evolution.}

However, our attention is paid rather to the more delicate
systematic trend in the [Na/Fe] vs. [Fe/H] relation; namely,
the ``upturn'' revealed by Edvardsson et al.'s (1993) analyses
based on the 6154/6161 lines, characterized by
slight upward increases on both sides of the broad minimum
around [Fe/H]~$\sim -0.2$. Carretta et al.'s (2000) reanalysis
on Edvardsson et al.'s (1993) data also reproduced this tendency,
as shown in figure 9 in their paper. Similarly, our reanalyzed
results on Edvardsson et al.'s (1993) data are essentially
the same as their original results (cf. small open circles
in figure 5c).

Supportive evidences for this trend have been reported from
successive analyses on independent observational materials.
Feltzing and Gustafsson (1998) confirmed in their
Na~{\sc i} 6154/6161 analyses of metal-rich stars that
[Na/Fe] progressively increases from
$\sim 0$ ([Fe/H] $\sim 0$) to  $\sim +0.2$ ([Fe/H] $\sim +0.4$)
(see the dots in figure 5c).
They also investigated its behavior in the kinematically different
stellar groups to see if there is any position-dependent effect in
the Galactic chemical evolution, though distinct differences were
not observed.
Also, Chen et al.'s (2000) LTE analyses of 6154/6161 lines
based on the BAO data of disk stars (common to this study)
suggested a weak minimum at [Fe/H] $\sim -0.2$ superposed
on the nearly flat [Na/Fe] ($\sim 0$) (cf. figure 8 therein)
consistent with that found by Edvardsson et al. (1993).

As mentioned in the previous subsection, the non-LTE corrections
on the Na~{\sc i} 6154/6161 lines usually used in sodium abundance
analyses of disk stars are so small that they are practically
negligible. Therefore, it is expected that our non-LTE analyses
on these orange lines of multiplet 5 reproduce essentially the same
results as obtained by Chen et al. (2000), which is actually
confirmed in figure 5c. We point out, however, that our non-LTE
[Na/Fe] values (on the BAO data) derived from other stronger lines
(5683/5688, 5890/5896, and 8195) also show signs of ``upturn'',
(tough not so clear as the case of 6154/6161 lines)
in spite of their inadequacy for abundance determinations,
as can be seen from figures 7a, b, and d.
Hence, we may state that this trend is firmly ascertained.

Considering the decline of the [Na/Fe] values for a decrease of [Fe/H]
both on the metal-poor side ([Fe/H] $\la -1$; cf. the next subsection)
and the metal-rich side ($0 \la$ [Fe/H]; cf. Feltzing, Gustafsson 1998),
the existence of such an ``upturn'' may indicate an extra production
of sodium at the cosmic metallicity of $-1 \la$ [Fe/H] $\la -0.5$.
However, any speculation had better be put off until the behaviors
of other elements closely related to the synthesis of Na, especially
Al and Mg, have been established.

According to our reanalysis results for Prochaska et al.'s
(2000) data (cf. diamonds in figure 7), the [Na/Fe] values of
thick disk stars do not appear to exhibit significantly distinct
differences from those of normal thin-disk stars, though our
mean non-LTE $\langle$[Na/Fe]$\rangle$ averaged over their 10 stars
are nearly zero [$-0.03 (\sigma = 0.06)$ and $+0.05 (\sigma = 0.04$) 
for 5683/5688 and 6154/6166 lines, respectively] and slightly smaller
than the value (0.087) they obtained, which may be due to the
non-LTE effect (mean non-LTE corrections are $-0.09$ and $-0.06$,
respectively).

\subsection{Halo Stars}

Now, we discuss the behavior of [Na/Fe] for metal-poor halo stars.
For simplicity, our discussion is confined to figure 7, the results of
our reanalyzes based on the data taken from the works after 1990,
which may presumably be more accurate and reliable than those
in figure 6 for the reasons mentioned in subsection 4.3.

One important feature that can be read from figure 7 is the systematic
decrease of [Na/Fe]; i.e., from [Na/Fe] $\sim 0$ at [Fe/H] $\sim -1$
down to [Na/Fe] $\sim -0.4$ (though a rather large diversity of 
$\sim \pm 0.2$) at [Fe/H] $\sim -2$. This means that we have reasonably 
reconfirmed the recent results of Baum\"uller et al. (1998) and 
Stephens (1999), as opposed to the previous belief until mid 1990's 
(cf. section 1).
It should be stressed that each of the different multiplet lines
[5683/5688 (Stephens et al. 1999; half-filled circles),
5890/5896 (McWilliam et al. 1995a; filled circles),
6154/6161 (Kraft et al. 1992; double circles)
and 8183/8195 (our OAO data; St. Andrew's crosses)] yield consistent
results with each other, in spite of the considerably different
extents of the non-LTE corrections. Thus, we may state that
this subsolar [Na/Fe] ratios in metal-poor stars of
$-3 \la$~[Fe/H]~$\la -1$ are firmly established,\footnote{
Some remark may be due regarding our non-LTE reanalysis results of
Pilachowski et al.'s (1996) 5890/5896 data (cf. filled triangles
in figure 7) for population II low-gravity giants
mostly having parameter values of $T_{\rm eff} \sim$ 4000--5000 K,
$\log g \sim$ 0--2, and $-2.5 \la$ [Fe/H] $\la -1.5$,
which do not reveal any such clear subsolar tendency
[$\langle$[Na/Fe]$\rangle$ = $0.00 (\pm 0.24)$ on the average]
unlike the other cases mentioned above, despite their original
LTE results showed a moderate subsolar trend
($\langle$[Na/Fe]$\rangle$ = $-0.17$; cf. Pilachowski et al. 1996).
We point out that some of their program stars are low temperature/gravity
K giants with parameters of $T_{\rm eff} \sim 4000$~K or $\log g \sim 0.0$,
which are outside the parameter grid of our non-LTE correction tables
and the applied non-LTE correction had to be evaluated by an extrapolation.
In these low temperature/gravity cases, it exceptionally happens that
the non-LTE effect tends to act as a line-weakening mechanism
(i.e., marginally positive non-LTE correction) due to the enhanced
$S_{\rm L}$  ($S_{\rm L}/B > 1$) in the line-forming region according to
our calculations, as mentioned in footnote 1 in subsection 2.2. Hence,
though appreciable positive non-LTE corrections are actually observed
for several stars in figure 7b (filled triangles), these are all such
low-gravity K giants for which this effect may have been rather erroneously
exaggerated by inaccurate extrapolations.
Also, because of the low-temperature nature of the Pilachowski et al.'s
(1996) program stars, the strengths of the 5890/5896 D lines are
considerably large (200--400 m$\rm\AA$) in spite the low metallicity
and comparatively large ambiguities are expected (e.g., equivalent-width
measurements, how to choose the damping parameter or the microturbulent
velocity). Consequently, our reanalysis results on their data may suffer
large uncertainties and should not be seriously taken.}
while previous LTE analyses using the non-LTE sensitive lines
(e.g., McWilliam et al. 1995b, who invoked 5890/5896 lines)
must have overestimated the sodium abundances.

Turning our attention to the further extreme metal-poor regime of
$-4 \la$ [Fe/H] $\la -3$, we see in figure 7 an interesting
trend of rising [Na/Fe] for a lowering of the metallicity,
again recovering [Na/Fe] $\sim 0$ at [Fe/H] $\sim -4$,
though a rather insufficient number of stars in this region
prevents us from making any definitive argument.
It is highly desired to increase the sample of those stars
toward establishing the behavior of [Na/Fe] in this ultra-low
metallicity region, which is very important for investigating
the history of the early-time Galaxy as well as for constructing
the realistic model of Galactic chemical evolution.

Finally, we compare the observed [Na/Fe] vs. [Fe/H] relations
obtained for the metal-deficient halo stars with representative
theoretical predictions calculated by using various chemical
evolution models of our Galaxy. As mentioned in section 1,
since the production of Na is dependent on the excess neutrons,
its yield should naturally be metallicity-dependent and increase
with time. As a matter of fact, the assumption of constant yield
results in a totally unrealistic [Na/Fe] vs. [Fe/H] relation with
a markedly negative $d$[Na/Fe]/$d$[Fe/H] gradient
(cf. the dashed line labeled ``GP(const)'' in figures 6 and 7, which
was taken from figure 7 of Goswami, Prantzos 2000).
It may be possible, therefore, to judge which modeling of Na production
(among several proposed chemical evolution calculations) represents
the actual situation best by comparing the predicted [Na/Fe] vs. [Fe/H]
relation with that observed for halo stars mentioned above.

For this purpose, theoretical predictions taken from three
recent representative works are depicted in figures 6 and 7; i.e.,
those of Timmes et al. (1995; solid lines), Samland (1998;
dashed-dotted line), and Goswami and Prantzos (2000; dashed lines).
We can draw from figure 7 the following conclusions.\\
--- Among these, Timmes et al.'s (1995) results, especially for the case 
of reduced Fe yield denoted as ``TWW (1/2 $\times$)'' (cf. figure 8a), 
appear to most satisfactorily reproduce the observed tendency of [Na/Fe] 
(i.e., broad/shallow dip-like feature with a minimum around [Fe/H] $\sim -2$) 
mentioned above.
Their calculations are based on Woosley and Weaver's (1995) yields,
simple dynamical model for the Galaxy with infall, a standard Salpeter
(1955) initial-mass function (IMF) and a quadratic Schmidt star-formation
rate. (Each line corresponds to the different choice of the Fe yield from
massive stars; i.e.,  that from the original Woosley and Weaver paper
and that reduced by a factor of 2; the latter is rather recommended by
Timmes et al. 1995).\\
--- Samland's (1998) model is based also on Woosley and
Weaver's (1995) yields. However, since he adjusted the metallicity-dependent
Na yield in his own way [cf. equation (9) therein] so as to fit the
observed trend of [Na/Fe] known at the time of mid-1990's (i.e., erroneously
overestimated to be [Na/Fe]$\sim 0$ even for halo stars), his prediction
does not fit the observed [Na/Fe] obtained in this study. \\
--- Goswami and Prantzos's (2000) calculation (labeled ``GP(var)''in
figure 7) explains the decline of [Na/Fe] from [Fe/H]$\sim -1$ to
[Fe/H]$\sim -2$ reasonably well. However, their prediction suggests an
ever-decreasing (or a plateau-like) [Na/Fe] toward the extremely
low metallicity regime ([Na/Fe]$\sim -4$), which apparently contradicts
the tendency we found. While their model is based on Woosley and Weaver's
(1995) metallicity-dependent yields (as was done by Timmes et al. 1995)
with a Fe yield reduced by a factor of 2,  they used a different IMF
(Kroupa et al. 1993) and a different halo model. From this point
of view, the dynamical model of the Galaxy and IMF adopted by Timmes et al.
(1995) might be preferable to those of Goswami and Prantzos (2000),
as far as the early Galaxy ([Fe/H] $\la -2$) is concerned.

\section{Conclusion}
\label{concl}
We carried out extensive non-LTE calculations on the
atomic model of neutral sodium and model atmospheres with a wide
range of parameters, for the purpose of non-LTE abundance
determinations based on eight representative Na~{\sc i} lines at
5683, 5688, 5890, 5896, 6154, 6161, 8183, and 8195~$\rm\AA$.

The non-LTE effect almost always acts as a line-strengthening
mechanism (thus the non-LTE abundance correction being negative).
Though its extent differs from line to line (from almost negligible
level of a few hundredths dex to a considerable amount of $\sim 0.5$ dex),
strong saturated lines (though not too strong to be damping-dominated)
tend to suffer a large non-LTE effect. Generally speaking, the lines
for which non-LTE corrections had better be taken into account in Na
abundance determinations are 5890/5896 and 8183/8195 lines, while
the 5683/5688 and (especially) 6154/6161 lines are comparatively
less sensitive to a non-LTE effect.

Our non-LTE reanalyzes of our BAO and OAO equivalent-width data
along with those taken from extensive literature revealed
the following conclusions:\\
--- (1) Regarding the disk stars with $-1 \la$ [Fe/H] $\la 0.4$,
we confirmed the existence of a delicate ``upturn'' feature
(i.e., broad/shallow dip around the minimum at [Fe/H] $\sim -0.2$)
superposed on the general tendency of [Na/Fe] $\sim 0$,
as has been reported by recent analyses on weak Na~{\sc i} 6154/6161
lines being inert to any non-LTE effect. Moreover, according to our
analyses on our BAO data, even the abundances derived from strong
(saturated or damping-dominated) lines, which are generally unsuitable
for abundance analyses, suggested this tendency.\\
--- (2) We found based on our analyses of recent data after 1990's
that the [Na/Fe] ratios for metal-poor halo stars show a ``subsolar''
behavior; i.e., [Na/Fe] decreases from $\sim 0$ (at [Fe/H]$\sim -1$)
to $\sim -0.4$ (at [Fe/H]$\sim -2$), while it appears to rise again
with a further decrease in metallicity toward recovering [Na/Fe]$\sim 0$
again at [Fe/H]$\sim -4$. It is evident that the previous suggestion of
almost solar Na-to-Fe ratio over all metallicity, which was actually
believed until mid 1990's, is attributed to the overestimation
of sodium abundances due to the neglect of non-LTE effects.
Among the representative theoretical models of Galactic chemical evolution,
that of Timmes et al. (1995) appears to reasonably reproduce this behavior.\\
--- (3) Given this observational evidence, the run of [Na/Fe] is in
rough accord with that of [Al/Fe] (cf. Baum\"uller, Gehren 1997),
which would be a sound consequence from a theoretical point of view,
since both are expected to show similar behaviors (see, e.g.,
subsection 5.8 of Samland 1998).
In order to further clarify the connection between the present results and
the abundances of such important Na-related elements in metal-poor stars,
we are now planning to carry out extensive non-LTE analyses on Al and Mg,
in the way similar to that adopted in this paper by using our own
as well as the literature data, the results of which will be reported
in our forthcoming papers.
\newline

We thank the staff members of OAO, especially H. Izumiura and S. Masuda,
for their kind support and suggestions concerning the operation and
maintenance of HIDES.
Thanks are also due to S. Honda, K. Sadakane, S. Sato, and K. Osada
for their helpful collaborations in our 2001--2002 observations at OAO.
This work was done within the framework of the China--Japan
collaboration project, ``Galactic Chemical Evolution through
Spectroscopic Analyses of Metal-Deficient Stars'' supported by
the Japan Society for the Promotion of Science (JSPS) and the Natural
Science Foundation of China (NSFC).

\end{document}